\documentclass[graybox]{svmult}

\usepackage{type1cm}        
\usepackage{makeidx}         
\usepackage{graphicx}        
\usepackage{multicol}        
\usepackage[bottom]{footmisc}
\usepackage{newtxtext}       %
\usepackage{newtxmath}       

\newcommand{\ti}{\text{\raisebox{-0.55ex}{$\;\widetilde{}$}\kern-.11em{$\shortmid\,$}}}
\newcommand{\intii}{\int_{-\infty}^{+\infty}}
\newcommand{\intoi}{\int_{0}^{\infty}}
\newcommand*\conj[1]{\overline{#1}}
\def\d{\text{d}}

\usepackage{musixtex}
\newcommand{\mus}{\protect\scalebox{0.55}{\trebleclef}}

\makeindex             

\usepackage{color}
\definecolor{crimson}{RGB}{220,20,60}
\definecolor{darkorchid}{rgb}{0.6, 0.2, 0.8}\definecolor{lightskyblue}{RGB}{135,206,250}
\definecolor{g}{rgb}{0.1,0.6,0.2}

\begin{document}

\title*{Quantifying the rationality of rhythmic signals} 
\author{A. Guillet \and A. Arneodo \and P. Argoul \and F. Argoul}
\institute{A. Guillet, A. Arneodo$^{\dag}$, F. Argoul \at CNRS, UMR5787, Laboratoire Ondes et Mati\`ere d'Aquitaine, Universit\'e de Bordeaux, France  \email{name@email.address}
\and P. Argoul \at MAST-EMGCU, Univ Gustave Eiffel, IFSTTAR,
F-77477 Marne-la-Vall\'{e}e, France \email{pierre.argoul@email.address}}
\maketitle

\abstract*{Rhythms and vibrations represent the quintessence of life, they are ubiquitous (systemic) in all living systems.  Recognising, unfolding these rhythms is paramount in medicine, for example in the physiology of the heart, lung, hearing, speech, brain, the cellular and molecular processes involved in biological clocks. The importance of  the commensurability of the frequencies in different rhythms has been thoroughly studied in music.   We define a log-frequency correlation measure on spectral densities that gives   the temporal evolution of the  distribution of frequency ratios (rational or irrational) in between  two signals,  using analytic wavelets. We illustrate these concepts on numerical signals (sums of sine functions) and voice recordings from the Voice-Icar-Federico II database. Finally, with a second correlation operation from two of these ratio distributions (a reference one, the other from the voices) we introduce another quantity that we call \emph{sonance}, measuring the ``harmony'' (rationality) of two voices sung together as a function of a pitch transposition.}  

\abstract{Rhythms and vibrations represent the quintessence of life, they are ubiquitous (systemic) in all living systems.  Recognising, unfolding these rhythms is paramount in medicine, for example in the physiology of the heart, lung, hearing, speech, brain, the cellular and molecular processes involved in biological clocks. The importance of  the commensurability of the frequencies in different rhythms has been thoroughly studied in music.   We define a log-frequency correlation measure on spectral densities that gives   the temporal evolution of the distribution of frequency ratios (rational or irrational) in between  two signals,  using analytic wavelets. We illustrate these concepts on numerical signals (sums of sine functions) and voice recordings from the Voice-Icar-Federico II database. Finally, with a second correlation operation from two of these distributions of ratios (a reference one, and the other one from the voices) we introduce another quantity that we call \emph{sonance}, measuring the ``harmony'' (rationality) of two voices sung together as a function of a pitch transposition.}  

\section{Introduction}
\label{sec:1}
Scientific approaches of natural systems have been revolutionized in the last part of the XXth century with the advent of miniaturized electronic and computer systems. Beyond their impressive beauty, it was offered to human beings to demonstrate that nature is constructed from multi-scale intertwined networks, (in time and in space) and that these networks are the field of highly complex  nonlinear dynamics (non linear and/or non stationary rhythms) \cite{jorgensen_encyclopedia_2008,ivanov_focus_2016,barabasi_network_2011}. 
Even though apparently distinct biological rhythms (endogenous and exogenous) have been recognised as universal features of all organisms (neural signals, heart, hormone secretion, metabolism,  tidal, circadian, lunar, seasonal, annual clocks, life cycle, ....) \cite{goldbeter_au_2018}, the variability of these rhythms and their spatio-temporal interplay is still considered as incidental or ignored. Despite the fact that we can concretely demonstrate that the frequencies of these rhythms pave more than 10 decades, still, time (and frequency) is considered as varying linearly in living systems. 
In particular the presence of strong nonlinearities can give  us greater  sensing  resolution  to  less  intense  stimuli. These mechanisms are ubiquitous across animal species and across all sensory  modalities. Interestingly, the mappings between  an external  stimuli  and  the internal  perception (psychophysical) of scales  and laws are rather logarithmic than linear. A simple and more commonly encountered example for the non-specialist is the  perception  and emission of acoustic vibrations (sounds) by living species, these processes occur in logarithmic scales in time and frequency domains \cite{oppenheim_human_2013}. It has also  been demonstrated experimentally that the cochlear filters of the inner ear are not spaced at linear frequency intervals but that their spacing is approximately logarithmic \cite{schnupp_auditory_2011}. 

The emission of sound (speech, songs) by human cord tract (larynx, pharynx, mouth) is a complex nonlinear process that combines both muscles and tissues with different temporal and spatial scales, and the entire autonomic and central nervous systems. 
In this study, we analyse human voice signals (a single note maintained for a few seconds) that characterise the physiology of the vocal organ (larynx-pharynx-mouth) in healthy and pathological situations. 
To compare different signals and their spectral composition, we define a log-frequency correlation measure on spectral densities that gives   the distribution of frequency ratios (rational or irrational numbers) between two signals. Using the wavelet transform formalism we extend this measure to a time-frequency correlation measure, that offers the possibility to  estimate the temporal variability of this log-frequency correlation. We introduce reference spectral expansions as sums of Dirac terms that resume the characteristic property of these voice signals (harmonics as integer multiples of a fundamental frequency). 
Finally, we  define a new integral  cross-correlation of the previously defined measure which quantifies the rationality of the rhythms of two compared signals. We call it \emph{sonance}, by analogy with the term consonance (resp. dissonance) that counts the perceived affinity or agreement  (resp. disagreement) between different sounds. 
We validate this method on numerical model and voice signals collected from different sources. 
The first section is this introduction. The second section describes the mathematical methodology for log-frequency correlations (or spectrum of frequency ratios) and its generalization to time-frequency expansions in terms of  analytic wavelet transforms.  The third section  illustrates these concepts on numerical signals (sums of sine functions) and voice recordings from the Voice-Icar-Federico II database, introduces the  \emph{sonance}  measure and illustrates it on the previously computed log-frequency correlation measures of voice signals.  Finally, we  leave the medical application of voice dysphonia diagnosis with the comparison of an untrained voice with a singer voice that have similar spectral envelopes. 

\section{Spectrum of frequency ratios. Formalism and time-frequency generalization}
\label{sec:2}
\subsection{Correlation functions for signal comparison}
Let us consider two signals $x$ and $y$ of finite total energy  $L^2(\mathbb{R})$ :
$\langle x,x\rangle<+\infty$  and $\langle y,y\rangle<+\infty$ where 
$\langle \cdot,\cdot\rangle$ is the ordinary inner product of $L^2(\mathbb{R})$ and $\conj{x}$ is the complex conjugate of $x$: 
\begin{equation}
\langle x,y\rangle = \intii \conj{x(u)} y(u) \d u  .
\label{eq:innerprod}
\end{equation}

The comparison of these two signals $x$ and $y$ is usually performed through a deterministic correlation function $R[x,y] (\xi)$ constructed from a time shift (translation) operator $\mathbf{T}_{\xi}$:
\begin{align} 
R[x,y] (\xi) = \langle x, \mathbf{T}_{\xi}y \rangle = \intii \conj{x(u)}y(u+\xi)\d u   .
\label{intercorrelation}
\end{align}
This definition, given for energy signals or square-integrable functions, can be extended to power signals.
Thus, for signals which can be described by sums of periodic functions (stochastic signals with finite power), the cross-correlation function reads:  \begin{align} 
C[x,y](\xi)=\lim_{T\rightarrow \infty} \frac{1}{2T}\int_{-T}^{T}  \conj{x(u)}y(u+\xi)\d u   , 
\label{intercorrelation_in_time2}
\end{align}

When $x=y$,  we get the auto-correlation function $C[x,x](\xi)$, that characterises the similarity between observations of a same signal as a function of the lag $\xi$ between them. The auto-correlation function is Hermitian: $C[x,x](-\xi)=\conj{C[x,x](\xi)}$. The absolute value of $C[x,x](\xi)$  is maximum at the origin, where the auto-correlation function is real, positive and equal to the power of the signal $x$. When the signal $x$ is real, this implies that  the auto-correlation function is real and even.

Note that when $u=t$, $t$ being the time variable, the function  $C[x,y](\xi)$ is the cross-correlation function commonly used for time signals  but $u$ could be replaced by any other type of variable, and in particular the frequency (or log-frequency) when comparing spectral signals, as will be discussed below. 

Translation-based correlation functions are very important for physics. They turn functions of a relative quantity (such as time or space position whose value depends on a translation from an arbitrary origin) into a function of an absolute quantity (such as time or space interval). However, the value of absolute quantities that have a physical dimension still depends on its comparison with an arbitrary standard: the physical unit. Since a scaling is involved, the unit plays the role of an arbitrary origin for the logarithm of these quantities. That is the reason why dilation-based correlation functions can be of interest for physics, as long as they compare functions of an absolute quantity: a new variable made of the ratio of two absolute quantities with the same physical unit neither depends on an origin nor on the unit; it is a pure proportion.

To extend the concept of correlation functions to absolute physical quantities, we need first to revisit the definition of the inner product. We make use of the logarithm to change from the translation-invariant group $(\mathbb{R},+)$ to the dilation-invariant one $(\mathbb{R}^+,\times)$. The change of variable $u = \log v$ applied on Eq.~\eqref{eq:innerprod} yields:
\begin{equation}
\langle X,Y\rangle = \int_{-\infty}^{\infty} \conj{X(u)}\, Y(u) \d u =  \intoi \conj{X(\log v)} Y(\log v) \d \log v   .
\label{eq:innerprod_log}
\end{equation}
The change from the function $X(u)$ and the measure $\d u$ to the function $X\circ \log(v)$ and the measure $\d \log v=\d v/v$ means, for numerical computations, that we replace linearly sampled functions by geometrically sampled ones (of positive variable). In the following, we choose to make explicit the composition with the logarithm in each function. 
The previous translation operator $\mathbf{T_{\xi}}$ is naturally replaced by a dilation operator $\mathbf{D}_q$:
\begin{align}
\mathbf{T}_{\log q}[X](\log v)=\mathbf{D}_q [X\circ\log](v) = X(\log (q v))   . \label{eq:dilationoper}
\end{align}

Combining Eqs~\eqref{eq:innerprod_log} and \eqref{eq:dilationoper}, we obtain from Eq.~\eqref{intercorrelation} a  similar correlation function adapted to geometrically sampled signals:
\begin{align} 
R[X,Y](\log q)=\int_{0}^{\infty} \conj{X(\log v)}Y(\log (q v))\d \log v   , 
\label{intercorrelation_in_logtime}
\end{align}
where $q$ is positive.
For functions $X$ and $Y$, the finite energy condition for the validity of this integral takes the form $\langle X,X\rangle < +\infty$. It can also be reformulated for finite power signals in a similar way as in Eq.~\eqref{intercorrelation_in_time2}. 

The dilation correlation function in Eq.~\eqref{intercorrelation_in_logtime} inherits the following symmetry and linearity properties from Eq.~\eqref{intercorrelation}:
\begin{align}
    R[Y,X](\log q)&=\conj{R[X,Y](-\log q)} \label{symmetry}  , \\
    R[X,Y+Z](\log q)&=R[X,Y](\log q)+R[X,Z](\log q)  . \label{linearity} 
    \end{align}

Note that the logarithm does not allow to study functions of a negative absolute quantity (for instance negative delays or frequencies), nor negative ratios $q<0$. 

\subsection{Spectrum of frequency ratios: a frequency ratio distribution}

For the application of interest here, the unfolding of rhythms from real signals (their spectral ``timbre''), we concentrate on ``geometric'' spectral densities that we define as real and positive functions $\mathcal{S}(\log f)\geq 0$ of the logarithm of the frequency. 
The log-frequency correlation function between two such densities  
\begin{align} \label{spectrumrelations}
R[\mathcal{S}_1,\mathcal{S}_2](\log q)= \intoi \mathcal{S}_1(\log f) \mathcal{S}_2(\log(q f))\d \log f  , 
\end{align}
captures all the spectral relations between frequency modes of $\mathcal{S}_1(\log f)$ and $\mathcal{S}_2(\log f)$.
$R[\mathcal{S}_1,\mathcal{S}_2](\log q)$ is positive and gives the distribution of frequency ratios $q$ of $\mathcal{S}_1(\log f)$ and $\mathcal{S}_2(\log f)$, hence the notation $R$ for ratio distribution. 
Similarly to standard correlation function of linearly sampled variables, the existence of this integral $R[\mathcal{S}_1,\mathcal{S}_2](\log q)$ requires that both distributions $\mathcal{S}_1(\log f) $ and $\mathcal{S}_2 (\log f)$ be square integrable with the geometric measure of $f$ (linear measure for $\log f$). 
Both the log-frequency distribution $\mathcal{S}(\log f)$ and the frequency ratio distribution $R[\mathcal{S}_1,\mathcal{S}_2] (\log q)$ can be normalised as probability density functions:
\begin{align}\label{normalisation}
\intoi R[\mathcal{S}_1,\mathcal{S}_2](\log q)\d \log q=\intoi \mathcal{S}_1(\log f)\d \log f \intoi \mathcal{S}_2(\log f)\d \log f =1 \;.
\end{align}

\begin{figure}
\begin{center}
\vspace{-0.3cm}
\includegraphics[scale=0.4]{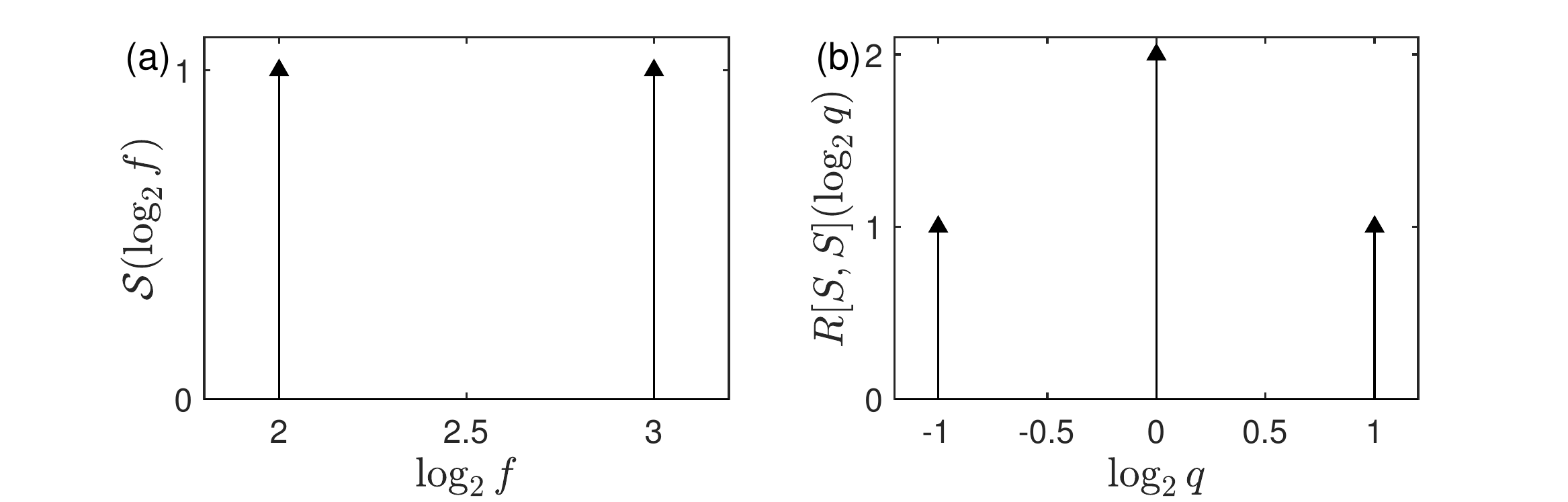} 
\caption{(a) Ideal distribution $\mathcal{S}(\log f)$ in log-frequencies of a doublet of Dirac deltas such that the highest frequency is twice the lowest. (b) Representation of the spectrum of self-relations $R[\mathcal{S},\mathcal{S}](\log q)$  in logarithmic scale (base 2). The peak of ratio $\log_2 q = 0$ represents the self-relation of each frequency peak, whereas the ratios $\log_2 q = -1, 1$ represent their cross-relations.}
\label{example1Rq}
\vspace{-0.8cm}
\end{center} 
\end{figure}

Frequency ratio distributions can be written in analytic form from spectral densities defined as isolated or sum of  Dirac  $\delta$ functions. For example, the two spectral densities $\mathcal{S}_j(\log f)=\delta(\log \frac{f}{f_j})$, $j=1,2$ have a single frequency ratio $\frac{f_2}{f_1}$, and give a frequency ratio distribution $R[\mathcal{S}_1,\mathcal{S}_2](\log q)=\delta (\log \frac{q f_1}{f_2})$. If we define $\mathcal{S}(\log f)$ as a doublet of Dirac deltas  $\mathcal{S}(\log f)=\mathcal{S}_1(\log f)+\mathcal{S}_2(\log f)$, from the linearity property Eq.~\eqref{linearity}, we can write the ratio distribution  $R[\mathcal{S},\mathcal{S}](\log q)=\delta (\log \frac{qf_1}{f_2})+2\delta (\log q)+\delta (\log \frac{qf_2}{f_1})$. This simple analytic case is illustrated in Fig.~\ref{example1Rq}, where we distinguish from  $R[\mathcal{S},\mathcal{S}](\log q)$ three peaks, corresponding to the frequency pairs: (4:4) and (8:8)  for $\log q = 0$, (4:8)  for $\log q = \log 2$, and  (8:4) for  $ \log q = -\log 2$. 

The log-frequency spectral distributions $\mathcal{S}(\log f)$ cannot be assimilated to  lin-frequency spectral densities defined the from Fourier transform of $s$:  $\hat{s}(f)=\!\intii \!s(t)e^{- 2\pi i ft}\d t$, because their computation from linear measures in time and frequency faces some difficulties. The main one is practical,  power spectral densities estimated with Fast Fourier Transform  (FFT) algorithms are sampled linearly, whereas the integral of Eq.~\eqref{spectrumrelations} requires a geometric frequency sampling. 
Re-sampling strategies  of the Fourier spectra have been proposed in the literature \cite{haines_logarithmic_1988}, and could be used for stationary signals, however they require greater memory size and are computer time-consuming. Importantly, 
 in the context of physiological signals which are often non-stationary, the extension of time-averaged spectral quantities to time-frequency distributions is mandatory. The wavelet transform answers to both issues, it provides not only  a time-frequency representation of the spectral quantities, but also allows a geometric sampling in frequency. Using time-frequency decompositions we can straightforwardly extend our definition of log-frequency ratio distributions Eq.~\eqref{spectrumrelations} to time-log-frequency ratio distributions for the analysis of non-stationary signals.

\subsection{Wavelet transform formalism} 
Time-frequency analysing tools based on the wavelet transform have been introduced in the second half of the twentieth century and  applied to many scientific domains for characterising and modelling non-stationary processes \cite{grossmann_decomposition_1984,kronland-martinet_analysis_1987,combes_wavelets_1989,delprat_asymptotic_1992,carmona_identification_1995,carmona_characterization_1997,carmona_practical_1998}. 
The wavelet transform of a finite energy signal $s(t) \in L^2(\mathbb{R})$ is defined as  its inner product with the shifted   copies of an analysing absolute integrable and finite energy  wavelet $\psi(t) \in L^1(\mathbb{R}) \cap L^2(\mathbb{R})$  \cite{kronland-martinet_analysis_1987,carmona_practical_1998,torresani_analyse_1995,flandrin_time-frequency_1998}:
\begin{equation}
\mathcal{W}_\psi^{(p)}[s](a,b) = \left< \psi_{a,b},s\right> = a^{-\frac{1}{p}}\int_{-\infty}^{+\infty} s(t)\conj{ \psi\left( \frac{t-b}{a}\right) }\,\d t ,
\label{eq:waveletdef_t}
\end{equation}
$b \in \mathbb{ R}$ and $a \in \mathbb{R}^{+}$ are the shift and scaling parameters. 
$\conj{\psi}$ is the complex conjugate of the analysing wavelet $\psi$, $p$ is a parameter which defines the normalisation of the wavelet. 

\noindent Two values of $p$ are usually found in the literature: $p=1$, corresponding to the $L^1(\mathbb{R})$ norm and $p=2$, corresponding to the $L^2(\mathbb{R})$ norm, respectively.

\noindent $p=1$,  often used for time-localized signals with different amplitudes, is appropriate 
when the magnitude of the modulus wavelet transform is wished to reflect the amplitude of the analysed signal $s(t)$. $p=2$  is appropriate 
when the modulus-squared wavelet transform is wished to reflect the energy of the analysed signal $s(t)$. 

In the frequency domain, the expression of the wavelet transform reads:
\begin{equation}
\mathcal{W}_\psi^{(p)}[s](a,b)= a^{1-\frac{1}{p}} \int_{-\infty}^{+\infty} \hat{s}(f) \;\conj{\hat{\psi}\!\left( a\!f\right) } e^{2 i \pi fb} \d f \; ,
\label{eq:waveletdef_f}
\end{equation}
where $\hat{s},\hat{\psi}$ denote the Fourier transforms of the signal and the wavelet.

\noindent This time-scale representation is quite suited for non-stationary signals since it localizes the analysis around time $b$ and operates a  band-pass filtering scaled by the parameter $a$. Importantly, $a$ can be sampled arbitrarily, in our case we will sample it geometrically. 
It is common practice to consider the scale $a$ as proportional to an inverse frequency $\frac{1}{f_{a}}$:
 \begin{equation}
     a=\frac{\f_{\psi}}{f_{a}} ,
     \label{a_fonction_de_la_frequence}
 \end{equation}
where $\f_{\psi}$ is a characteristic frequency of the mother wavelet $\psi$.
 Three meaningful frequencies are classically used for $\f_{\psi}$ \cite{lilly_higher_2009}: the peak frequency $f_{\psi}^{0}$ where the frequency domain mother wavelet magnitude $\left| \hat{\psi}(f) \right|$ is maximum, the energy (norm 2) frequency $\f_{\psi}^{\ast}$ which is the mean of $\left| \hat{\psi}(f) \right|^{2}$ and the norm 1 frequency $\check{f}_{\psi}$, that can be interpreted as an instantaneous frequency for progressive wavelets.
 An asymmetry in the frequency domain of the mother wavelet leads to distinct values for the previous frequencies $f_{\psi}$.
 
 \noindent For the computation of the log-frequency correlation functions, the expression $\mathcal{W}_\psi[s](f_{\psi}/f_a,b)$ for the wavelet transform given in Eq.~(\ref{eq:waveletdef_f}) can be turned to a time-frequency analysis by using Eq.~(\ref{a_fonction_de_la_frequence}) for a given characteristic frequency $f_{\psi}$:
\begin{equation}
\mathcal{W}_\psi^{(p)}[s]\left(\frac{f_{\psi}}{f_a},b\right)= a^{1-\frac{1}{p}}\, \int_{-\infty}^{+\infty} \hat{s}(f) \;\conj{\hat{\psi}\!\left( \frac{f_{\psi}}{f_a}\!f\right) } e^{2 i \pi fb} \d f \; .
\label{eq:waveletdef_f2}
\end{equation}

\noindent For our applications to physiological signals,
 the Banach space $L^1(\mathbb{R},\d t)$ norm corresponding to $p=1$ will be preferred for the wavelet transform definition.
 The main reason is due to the fact that when rescaling time in the input signal as $s\left( \frac{t}{\rho}\right)$, with $\rho >0$,
 both the time and the scale of the wavelet transform are rescaled, but without changing its magnitude. Thus as the Fourier transform of $s\left( \frac{t}{\rho}\right)$ is: $\rho\,\widehat{s}(\rho\,f)$,  
 Eq. (\ref{eq:waveletdef_f}) when $p=1$ leads to $\mathcal{W}_{\psi}^{(1)}\left[s\right](\frac{a}{\rho},\frac{b}{\rho})$. The (1) is dropped in the following. 

\noindent The peak frequency $f_{\psi}^{0}$ will be then adopted for the characteristic frequency $\f_{\psi}$ in Eqs \eqref{a_fonction_de_la_frequence}, \eqref{eq:waveletdef_f2}.

\noindent The admissibility condition for an analysing wavelet $\psi \in L^1(\mathbb{R}) \cap L^2(\mathbb{R})$ establishes that the number 
\begin{align}
c_{\psi}=\int_{0}^{+\infty} |\hat{\psi}(u)|^2 \;  \frac{\d u}{u} 
\end{align}
must be finite,   nonzero and independent of $f \in \mathbb{R}^+$. 
If this admissibility condition is fulfilled, then every $s \in L^2(\mathbb{R})$
can be reconstructed from  the convergent integral:
\begin{align}\label{CWTreconstruction}
   s(t) = \frac{1}{c_{\psi}}\intii\int_{-\infty}^{\infty} \mathcal{W}_\psi[s]\left(a,b\right)  \psi\left( \frac{t-b}{a}\right)  \frac{\d a}{|a|} \d b \;.
\end{align}

\subsubsection{Time and frequency window for the analysing wavelet}

The time-frequency window can be computed from the expression of the analysing wavelet $\psi$, assuming that $\psi$ and $\hat{\psi}$ verify $t \psi(t) \in L^2$ and $f \hat{\psi} (f) \in L^2(\mathbb{R})$  \cite{chui_introduction_1992}.
If the center and the radius (with the norm 2) of the window function $\psi$ are respectively $t_{\psi}^*$ and $\Delta_{\psi}$, $\psi((t-b)/a)$ is a window function with center $b+at_{\psi}^*$ and radius equal to $a \Delta_{\psi}$:
\begin{equation}
[ b+at_{\psi}^* - a\Delta_{\psi}, b+at_{\psi}^* + a \Delta_{\psi}] \; .
\end{equation}
This windows narrows (respectively widens) for small (resp. large) values of $a$. In the frequency domain, the window of $\hat{\psi}$ is defined similarly, assuming that the center and width of $\hat{\psi}$ are $f_{\psi}^*$ and $\Delta_{\hat{\psi}}$, $\psi (af)$ is centered around $f_{\psi}^*/a$ and has a radius $\Delta_{\hat{\psi}}/a$:
\begin{equation}
\left[ \frac{f_{\psi}^*}{a} -  \frac{1}{a} \Delta_{\hat{\psi}},  \frac{f_{\psi}^*}{a} +  \frac{1}{a} \Delta_{\hat{\psi}}\right] \; .
\end{equation}
In the following discussion, the center $f_{\psi}^*$ of $\hat{\psi}$ is assumed to be positive. There are different ways of defining the wavelet resolution, called the quality factor of the wavelet. A first definition, given in~\cite{le_continuous_2004}, uses the bandwidth and the norm 2 frequency as follows:
\begin{equation}
Q^{*} = \frac{f_{\psi}^*/a}{2 \Delta_{\hat{\psi}}/a} = \frac{f_{\psi}^*}{2 \Delta_{\hat{\psi}}} \; ,
\end{equation}
which is independent of the scale parameter $a$.
Alternatively, we could also use the full width at half maximum height of $|\hat{\psi}(f)|^2$ instead of $\Delta_{\hat{\psi}}$.
We thus define another quality factor, $\tilde{Q}$, such as
\begin{equation}
\tilde{Q} = \frac{f_\psi^0}{|f_2-f_1|}
\end{equation}
where $|\hat\psi(f_1)|^2 = |\hat\psi(f_2)|^2 = |\hat\psi(f_\psi^0)|^2/2$ and $f_1 < f_\psi^0 < f_2$.
This factor is usually computed to characterise the qualitative damping behavior of simple damped oscillators \cite{rocard_dynamique_1943}.

The choice of the quality factor is essential to obtain an adapted time-frequency resolution and consequently a ``good'' analysis of the processed signals. The authors in~\cite{erlicher_modal_2007} propose three bounds to obtain a range of acceptable values. When the signal is composed of several frequency components, the proximity of their characteristic frequencies provides a lower bound. The exponential decay rate of the amplitude imposes another upper bound. Eventually, the length of the signal determines yet another upper bound.

\subsubsection{Choice of the analysing wavelet: the Grossmann wavelet}

\noindent In the absence of a suitable unifying theory for wavelet behaviors, the
choice of a particular wavelet for a particular problem may often
appear arbitrary. For rhythmic signals, complex analytic analysing wavelets are preferred, leading to:
$\hat{\psi}(f)=0$, $\forall f \le 0$. In that case,  the measure appears naturally in these integrals (Eq.~\eqref{CWTreconstruction}), as in Eqs~\eqref{eq:innerprod_log} and \eqref{intercorrelation_in_logtime},  because the analysing wavelet is scale invariant (under dilations).

In the following, we choose a single-parameter progressive wavelet, introduced for the decomposition of Hardy functions by Grossmann and Morlet~\cite{grossmann_decomposition_1984}:
\begin{align}
\hat{\psi}_Q(f)=\left\{\begin{array}{ll}
\psi_0 e^{-\frac{1}{2}\left(Q\log \frac{f}{f_0}\right)^2} \quad &\forall f>0 \;;\\
0 &\forall f\leq 0 \;,
\end{array}\right.
\label{PsiQ_wavelet}
\end{align}
of peak frequency $f_\psi^0=f_0$, for which the maximum value is $\psi_0$. This wavelet is symmetric in log-frequencies about $\log f_0$. The other characteristic frequencies are $f_\psi^*=f_0e^{\frac{3}{4Q^2}}$ and $\check{f}_\psi=f_0e^{\frac{3}{2Q^2}}$. The Grossmann wavelet is also centered in time ($t_\psi^\ast=0$), with a width (radius in norm 2):
\begin{equation}
\Delta_\psi=\frac{\sqrt{1+2Q^2}}{4\pi f_0}\;.
\end{equation}
Both previously defined quality factors depend on $Q$ only:
\begin{align}\label{Qstar}
Q^*&=\tfrac{1}{2}\left(e^{\frac{1}{2Q^2}}-1\right)^{-\frac{1}{2}}   , \\ 
\tilde{Q}&=\left(2\sinh{\frac{\sqrt{\log 2}}{Q}}\right)^{-1}   .
\label{quality_factors}
\end{align}
When $Q$ is large enough, the leading term in the expansions gives $Q^*\simeq Q/\sqrt{2}$ and $\tilde{Q}\simeq Q/\sqrt{\log 2}$ respectively, followed by a term of order $\frac{1}{Q}$. Consequently, we will refer to the parameter $Q$ as the quality factor for this wavelet.

When choosing the value $\psi_0^2=\frac{Q}{\sqrt{\pi}}$, the admissibility constant is one and $|\hat{\psi}(f)|^2$ can be considered as a probability density function in log-frequencies:
\begin{align}
c_{\psi}=\intoi |\hat{\psi}_Q(f)|^2 \d \log f = 1 \;.
\end{align}

\begin{figure}
\begin{center}
\vspace{-0.3cm}
\includegraphics[scale=0.26]{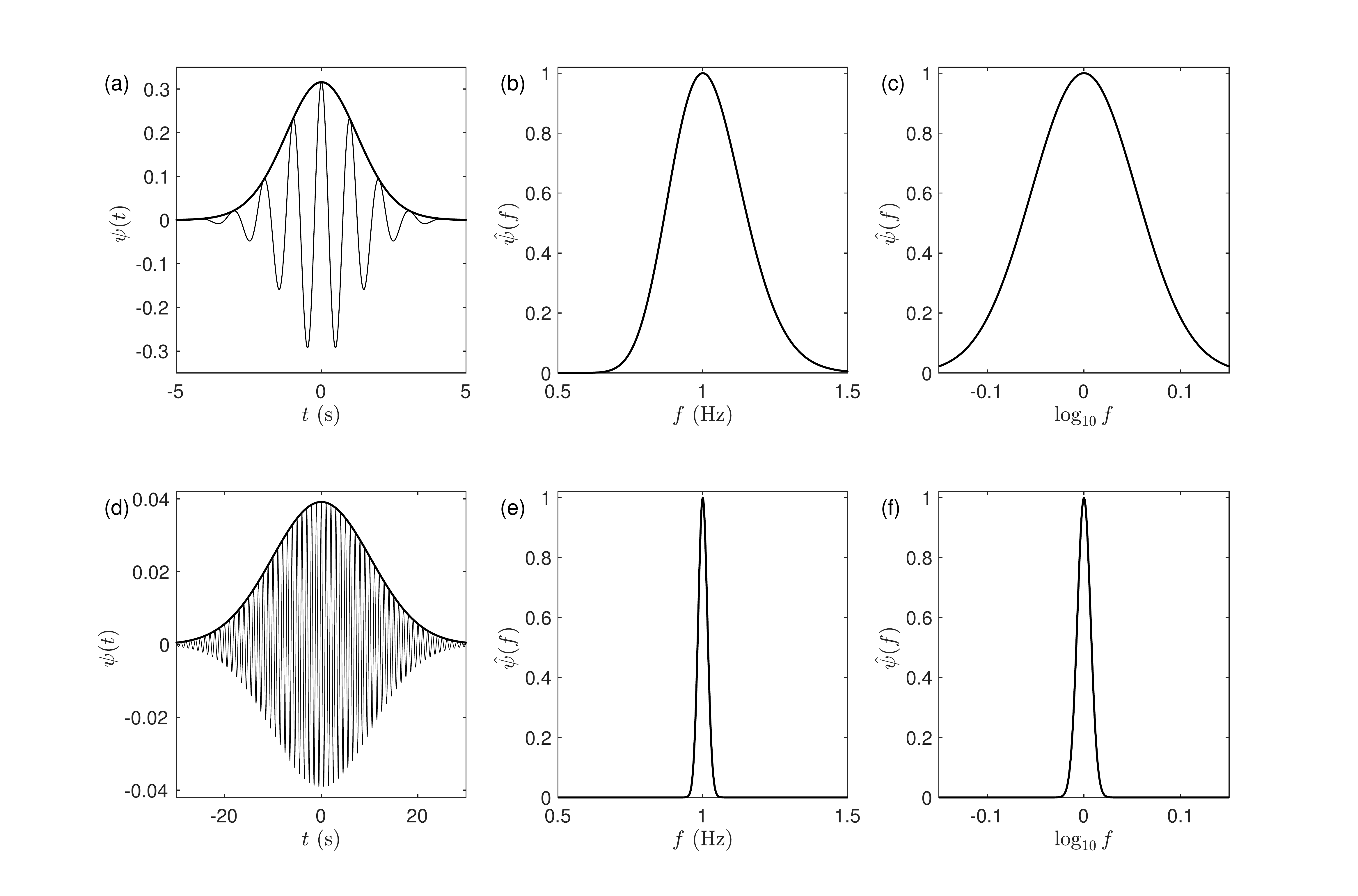}
\vspace{-0.6cm}
\caption{Grossmann analysing wavelet (log-normal in frequency). $ \psi (t)$ is computed by inverse Fourier transform of $ \hat{\psi}_Q(f)$ (Eq.~\eqref{PsiQ_wavelet}).   (a)  $\psi_Q (t)$.  (b)  $ \hat{\psi}_Q (f)$  in linear-frequency scale. (c)  $ \hat{\psi}_Q (f)$  in a (base 10) logarithmic frequency scale. (a,b,c) are computed  for $Q=8$. (d,e,f) Same as (a,b,c) for $Q=64$.  In (a) and (d) $|\psi_Q (t)|$ (respectively $\Re\left\{\psi_Q (t)\right\}$) are plotted in thick (resp. light) black lines. }
\label{lognormal_wavelet}
\vspace{-0.8cm}
\end{center} 
\end{figure}

In Fig.~\ref{lognormal_wavelet}, we plot the Grossmann wavelet for
two values of $Q$, respectively $Q=8$ (top plots) and $Q=64$ (bottom plots). For larger $Q$ values, the number of oscillations of $\Re\left\{\psi_Q (t)\right\}$ and its width increases whereas $\hat{\psi}_Q (f)$ narrows. We can observe in Fig.~\ref{lognormal_wavelet}(c,f) that the wavelet in the log-frequency domain is symmetric around $\log f_0 = 0$ whereas it is asymmetric around $f_0=1$ in linear frequencies (Fig.~\ref{lognormal_wavelet}(b,e)).
 An important aspect of the oscillating progressive wavelets is how many oscillations are fitting inside their time-window~\cite{le_continuous_2004,lilly_higher_2009}. This number of oscillations determines the acuteness of the local frequency detection of a given rhythm and is of order $Q$. If this number is too large, the wavelet averages over too much oscillations and cannot provide a correct estimation. Conversely, if the number of oscillations is insufficient (less that $\sim 3$) the detection of a local rhythm will not be possible. The choice of this parameter is particularly important if the signal presents sharp transitions or close frequencies, as will be illustrated in the following figures.

 The authors in~\cite{lilly_higher_2009} showed that the Grossmann wavelet can be seen as a scaling limit of a general family of progressive wavelets with two parameters, the Morse wavelet~\cite{daubechies_time-frequency_1988-1,morse_diatomic_1929-1,daubechies_wiener_1987,olhede_generalized_2002,lilly_analytic_2010,lilly_generalized_2012,lilly_element_2017}. The Cauchy-Paul wavelet, intensively used in quantum mechanics and in the context of analytic functions
\cite{paul_wavelets_1992}, as well as the analytic version of the derivative of Gaussian wavelet or the Airy wavelet all belong to the Morse family.

\subsection{Extension of frequency ratio distributions to time-frequency ratio distributions}
From the Grossmann progressive wavelet transform defined in the previous section, we define a time-frequency distribution for non-stationary signals:
\begin{align}\label{tfdist}
\mathcal{S}^{(Q)}(\log (f_a),b)=\left|\mathcal{W}_{\psi_Q}[s]\left(\frac{f_0}{f_a},b\right)\right|^2  .
\end{align}
Note that the integral of the wavelet transform definition in Eq.~\eqref{eq:waveletdef_f} is sampled linearly in $f$, but that the values of the frequencies $f_a$ (or scale $a$) can be chosen arbitrarily, for our purpose we will select them geometrically distributed.  In the following, $b=t$ and $f_a = f$ are considered as time and frequency parameters, which simplifies the notation of $S^{(Q)}(\log f, t)$. 
This distribution is computed for strictly positive values of $f$ and we can extend the definition of the cross-correlation function to  time-frequency distributions:
\begin{align} 
R[\mathcal{S}^{(Q)}_1,\mathcal{S}^{(Q)}_2](\log q,t) &= \intoi \mathcal{S}^{(Q)}_1(\log f,t) \mathcal{S}^{(Q)}_2(\log(q f),t)\d \log f  \label{spectrumrelations2} \\ & = \intoi \left|\mathcal{W}_{\psi_Q}[s_1]\left(\frac{f_0}{f},t\right)\right|^2 \left|\mathcal{W}_{\psi_Q}[s_2]\left(\frac{q f_0}{f},t\right)\right|^2 \d \log f   . \label{spectrumrelations3}
\end{align}
The log-frequency autocorrelation function is defined as $R[\mathcal{S}^{(Q)},\mathcal{S}^{(Q)}](\log q,t)$.
The temporal mean of  $\mathcal{S}^{(Q)} (\log f, t)$: $\langle\mathcal{S}^{(Q)}\rangle_t\!(\log f)$ that can be seen as a power spectral density based on the wavelet transform $\mathcal{W}_{\psi_Q}[s]$.

\subsubsection{Computation of the log-frequency correlation function}

Using the convolution theorem, $R[\mathcal{S}^{(Q)}_1,\mathcal{S}^{(Q)}_2] $ can be computed quite efficiently using the fast Fourier transform (FFT) several times (that discretizes the Fourier transform here denoted $\mathcal{F}$): on a first step with respect to the time variable the signal (noted $\mathcal{F}$), and on a second step with respect to the log-frequency variable  (noted $\mathcal{F}_{\log f})$ and the computation step is  an inverse FFT in log-frequency space (noted $\mathcal{F}_{\log f}^{-1}$). 
\begin{align}\label{Rcomp_numeric}
R[\mathcal{S}^{(Q)}_1,\mathcal{S}^{(Q)}_2](\log q,t)&= \mathcal{F}_{\log f}^{-1}\left[ \conj{\mathcal{F}_{\log f}\left[|W_{\psi_{Q}}[s_1](.,t)|^2\right]}\mathcal{F}_{\log f}\left[|W_{\psi_{Q}}[s_2](.,t)|^2\right]\right](\log q) \\
\text{where } W_{\psi_Q}[s](f,t)&=\mathcal{F}^{-1}\left[\conj{\hat{\psi}}_Q\left(\frac{f'}{f}\right)\mathcal{F}[s](f')\right](t) \; .
\end{align}

This supposes that the frequency $f$ (or scale $a$) parameter of the CWT is sampled geometrically. The slowest operations consist in matrix multiplications. The fact that the second step requires Fourier transforms of the distributions $\mathcal{S}$ on log-frequency scale implies that the computed range of log-frequency values is enlarged, and padded with zeros to avoid extra-ratios arising from the FFT computation by an artificial periodisation of the $\mathcal{S}$ distribution.

  \begin{figure}
\begin{center}
\includegraphics[scale=0.25]{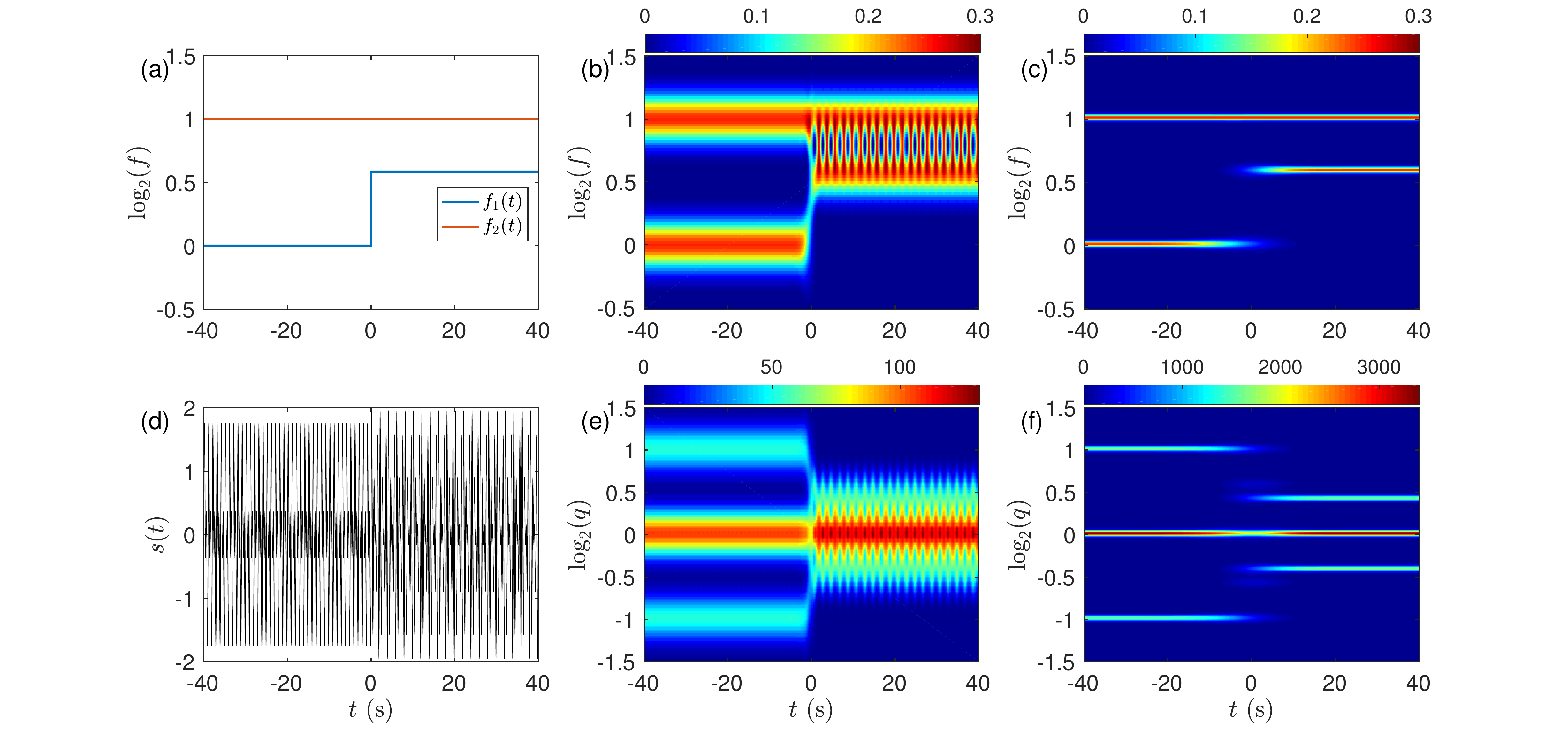}
\caption{Analysis of a model signal  defined as the sum of two sine functions $s(t)=\sin(2\pi f_1(t) t)+\sin(2\pi f_2 t)$, with $f_2 = 2$ constant, and $f_1(t)  = H(-t)+ \frac{3}{2} H(t)$ with the Heaviside step function. (a) Plot of the frequencies $f_1(t)$ and $f_2(t)$ in (base 2) logarithmic scale. (b) $ \mathcal{S}^{(8)} (\log f,t) $, computed for $Q=8$. (c) $\mathcal{S}^{(64)}( \log f,t)$, computed for $Q=64$. (d) Temporal signal $s(t)$ in the time window [-40s, 40s]. (e) $R[\mathcal{S}^{(8)},\mathcal{S}^{(8)}](\log q,t)$. (f) $R[\mathcal{S}^{(64)},\mathcal{S}^{(64)}](\log q,t)$.  $R[\mathcal{S}^{(Q)},\mathcal{S}^{(Q)}] (\log q,t)$ is defined in Eq.~\eqref{spectrumrelations2}. }
\label{step_frequency}
\vspace{-0.6cm}
\end{center} 
\end{figure}

\section{Computation of log-frequency distributions from numerical and real signals}
\subsection{Model signals constructed from sine functions}
\begin{figure}
\begin{center}
\vspace{-0.3cm}
\includegraphics[scale=0.25]{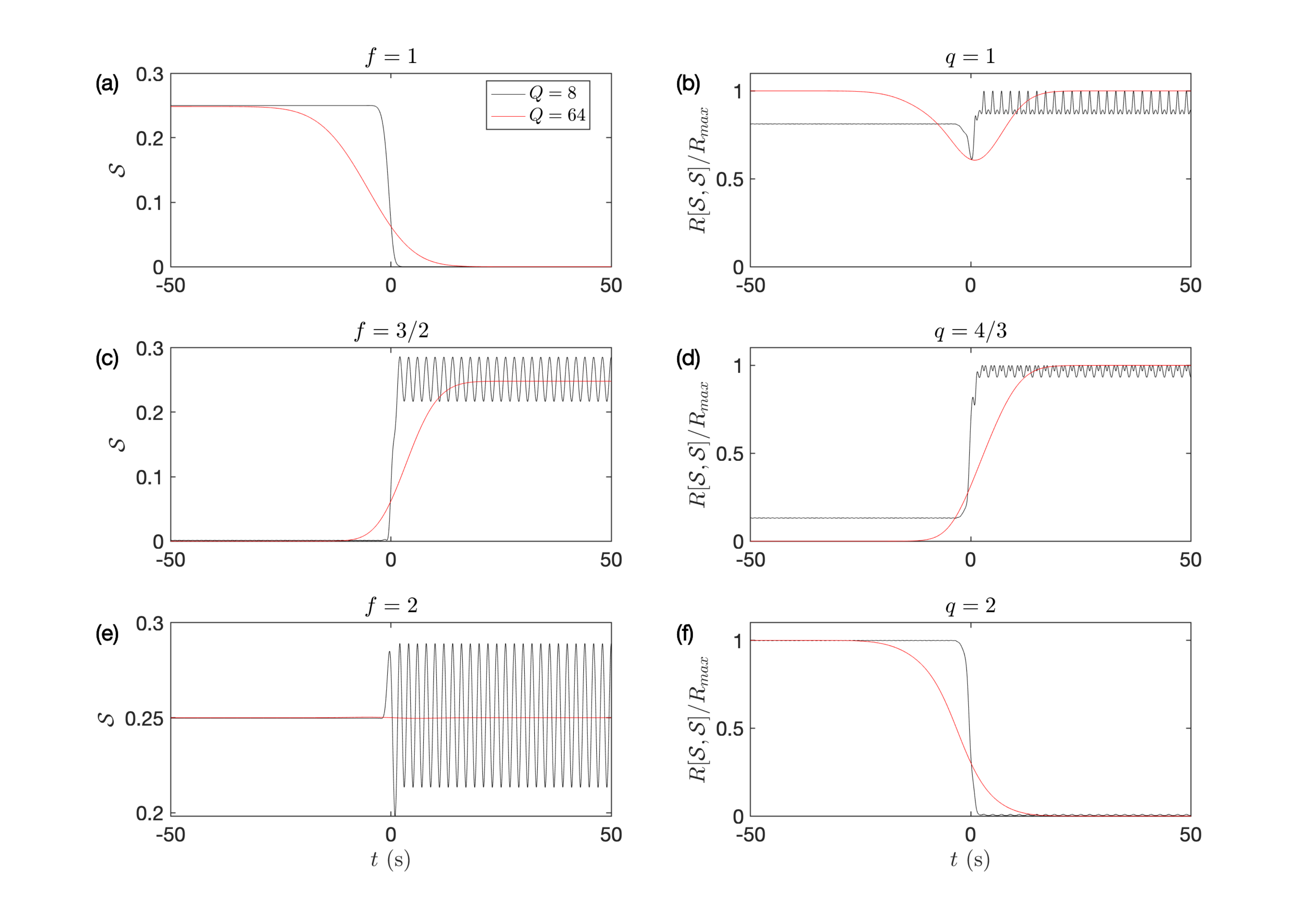}
\vspace{-0.3cm}
\caption{ (a, c, e) Sections of the log-frequency spectral distributions $\mathcal{S}^{(8,64)} (\log f_i, t)$ selected from Fig.~\ref{step_frequency}(b,c) for the three frequencies $f_1 = 1$, $f_2 = 3/2$ and $f_3 = 2$~Hz, and the same values of the quality factor $Q$ (8 and 64). (b, d, f) Sections of $R[\mathcal{S}^{(Q)},\mathcal{S}^{(Q)}](\log q_i,t)$ selected from Fig.~\ref{step_frequency}(e,f) for three values of $q$: $q_1 = 1$, $q_2 = 4/3$, $q_3 =2$, corresponding to local maxima of $R[\mathcal{S}^{(Q)},\mathcal{S}^{(Q)}]$. For each selected $q$ value, $R[\mathcal{S}^{(Q)},\mathcal{S}^{(Q)}]$ was scaled by its maximum in the time interval.}
\label{step_frequency_2}
\vspace{-0.8cm}
\end{center} 
\end{figure}

In Fig.~\ref{step_frequency}, we construct an artificial non-stationary signal from the sum of two sine functions: $s(t)=\sin(\phi_1(t))+\sin(\phi_2(t))$, with $\phi_2(t) = 4\pi t$ linear in time, and $\phi_1(t)  = 2\pi t H(-t)+ 3\pi t H(t)$ with the Heaviside step function $H$, and we compare the wavelet transform analysis for the two quality factors $Q=8$ and $Q=64$. 
With this signal we estimate a lower bound of $Q$ that is suitable for a frequency discrimination according to~\cite{erlicher_modal_2007}: for $t<0$, $Q\gtrsim 10$ and for $t>0$, $Q\gtrsim14$. Moreover, the signal length gives the constraint $Q\lesssim 285$. 
For $t<0$, the signal possesses two frequencies, highlighted on the colour-coded image of $ \mathcal{S}^{(Q)} ( \log f,t) $ by two horizontal bands ($f_1=1$ and $f_2=2$), the width of which depends on the quality factor $Q$, ($Q=8$ near the lower acceptable $Q$ bound in Fig.~\ref{step_frequency}(b) and $Q=64$ in Fig.~\ref{step_frequency}(c)). 
For $t>0$, we can again recognise the two bands  $f_1$ and $f_2$, and, as for $t<0$, their narrowing for the larger $Q$ values. 
The transition zone of this two bands, below and above $t=0$, needs to be discussed. Fig.~\ref{step_frequency_2} highlights this transition  with sections of $\mathcal{S}^{(Q)}( \log f,t)$ performed for remarkable values of $f$; 1, 3/2 and 2.  From the sections of Fig.~\ref{step_frequency_2}(a), we estimate  the width of this transition $\sim 4.8$s for $Q=8$, and $\sim 39$s for $Q=64$.   Another interesting phenomenon emerges in the $t>0$ regime, where the two frequency bands become closer. A low-frequency modulation of the wavelet transform squared modulus in the intermediate frequency range $[f_1, f_2]$ with 
 period $2$s appears, corresponding to frequency  $f_m = f_2 - f_1$ (0.5Hz in this example).  The matrix of the wavelet transform modulus is not simply the superimposition of the wavelet transform squared moduli of the sine alone, $|\mathcal{W}_{\psi_{Q}}[s_1+s_2] (a,b) |^2\ne |\mathcal{W}_{\psi_{Q}}[s_1] (a,b) |^2+|\mathcal{W}_{\psi_{Q}}[s_2] (a,b) |^2$
 , but extra terms such as $2\hat{\psi} (af_1) \hat{\psi} (af_2) \cos(2\pi (f_2-f_1) b)$ are also involved and are not negligible when $f_1$ and $f_2$ become too close (which is the case in Fig.~\ref{step_frequency}(b)). This effect disappears quite completely for larger $Q$ values because the product $\hat{\psi} (af_1) \hat{\psi} (af_2)$ vanishes. We conclude that the choice of $Q$ is a compromise between two objectives, (i) discriminating close frequencies (in which case larger $Q$ values  will be preferred), (ii) affording a correct temporal resolution for the detection of steep frequency changes (in which case smaller $Q$ values will be more efficient). 

Fig.~\ref{step_frequency}(e,f) shows the corresponding colour-coded maps  $R[\mathcal{S}^{(Q)},\mathcal{S}^{(Q)}] (\log q, t)$ for the same signal and the same values of $Q$ (8 and 64).  We recognise for $t<0$ three horizontal bands of constant $q$, corresponding respectively to frequency ratios $ q = 1/2, 1, 2$. The intensity of the middle band ($q=1$) is more contrasted ($\times 2$) because it corresponds to the sum of self-relations ($f_1$:$f_1$) and ($f_2$:$f_2$). The two symmetric weaker bands correspond to cross-frequency ratios ($f_1$:$f_2$) and ($f_2$:$f_1$). For $t>0$, the three bands become closer, and similarly to the maps of $\mathcal{S}^{(Q)}$ in Fig.~\ref{step_frequency}(b), a slow temporal modulation of $R[\mathcal{S}^{(Q)},\mathcal{S}^{(Q)}] (\log q,t)$ superimposes to the bands, due to coupling terms in the wavelet transform modulus. As expected, and similarly to what was observed on $\mathcal{S}^{(Q)} (\log f,t)$ maps, increasing $Q$ from 8 to 64 produces a strong narrowing of the bands and a strong reduction of the low-frequency modulation. The sections at fixed $q$ of these $R[\mathcal{S}^{(Q)},\mathcal{S}^{(Q)}] (\log q, t)$ maps are shown in Fig.~\ref{step_frequency_2} to highlight similarly the transition zone around $t=0$, its widening for larger $Q$ values, and the slow temporal modulations observed for $Q=8$. Due to the use of the Grossmann wavelet, sections at fixed $t$ of both $\mathcal{S}^{(Q)}(\log f, t)$ and $R[\mathcal{S}^{(Q)},\mathcal{S}^{(Q)}] (\log q, t)$ are Gaussian of widths $(\sqrt{2}Q
)^{-1}$ and $Q^{-1}$ respectively when the bands are not interfering (independent of $t$).

\begin{figure}
\begin{center}
\hspace*{-0.9cm}\includegraphics[scale=0.235]{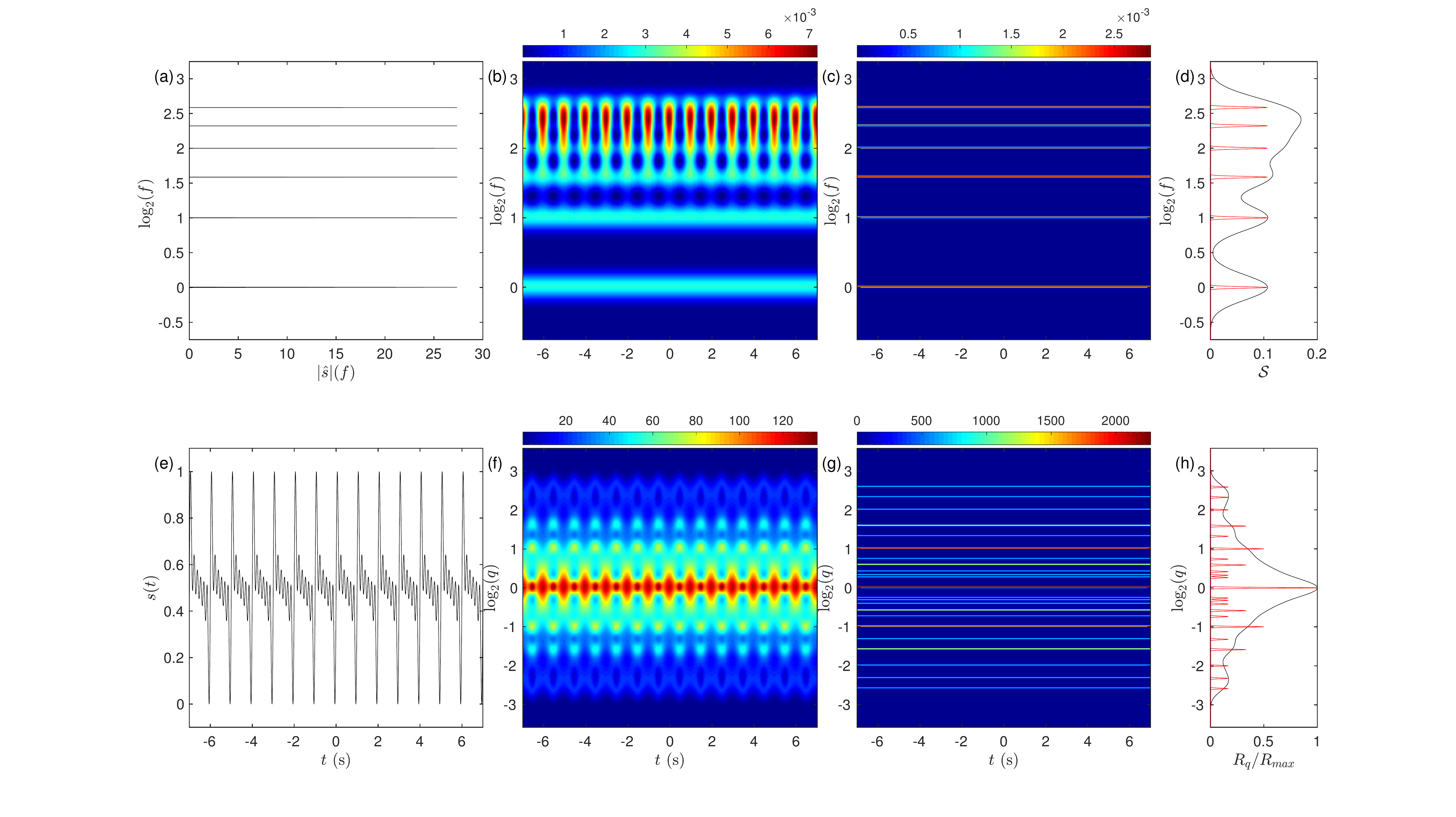}
\vspace{-0.5cm}
\caption{Analysis of a model signal  defined as the sum of 6 sine functions $s(t)=\sum_{1}^{6}\sin(2\pi f_i t)$, with $f_i = if_1$ constant. (a) Fourier spectra of $s$: $|\hat{s}| (f)$ on which the two axes have been inverted.  (b) $ \mathcal{S}^{(8)} ( \log f,t) $, computed for $Q=8$. (c) $\mathcal{S}^{(128)}( \log f,t)$, computed for $Q=128$. (d) $\mathcal{S}^{(Q)} ( \log f,t=0)$. (e) Temporal signal $s(t)$ in the time window [-6.5s, 6.5s]. (f) $R[\mathcal{S}^{(8)},\mathcal{S}^{(8)}](\log q,t)$. (g) $R[\mathcal{S}^{(128)},\mathcal{S}^{(128)}](\log q,t)$. (h) $R[\mathcal{S}^{(Q)},\mathcal{S}^{(Q)}] (\log q,t=0)$. In (d) and (h) the plots for $Q=8$ (resp. $128$) are coloured in black (resp. red). $R[\mathcal{S}^{(Q)},\mathcal{S}^{(Q)}] (\log q,t)$ is defined in Eq.~\eqref{spectrumrelations2}. }
\label{sum_sin_model}
\vspace{-0.8cm}
\end{center} 
\end{figure}

Another family of model signals (Fig.~\ref{sum_sin_model}(e)), particularly interesting with respect to the applications to voice signals, is defined as the sum of sine functions $\sum_{i=1}^{n} \sin (2\pi f_i t)$ with $f_i = i f_1$, $i$ positive integer. We use here a simple model with discrete and constant frequency components, that does not pretend to account neither for the intrinsic randomness nor the nonstationarity of physiologic signals. To improve the matching of model equations with real signals, we suggest two recent works based on the implementation of stationarity-breaking operators on Gaussian stationary random signals \cite{meynard-spectral_2018,omer-time-frequency_2017}.  
 In Fig.~\ref{sum_sin_model}, we take $n=6$ and perform the same time-frequency analysis with a Grossmann analysing wavelet with two values of $Q$, respectively $Q=8$ (b,c) and $Q=128$ (f,g). We note again that the larger $Q$, the finer and distinguishable the peaks of both $\mathcal{S}^{(Q)}( \log f,t)$ and $R[\mathcal{S}^{(Q)},\mathcal{S}^{(Q)}] (\log q,t)$. The already noticed low-frequency modulations in the previous example again appear in this example (Fig.~\ref{sum_sin_model}(b,f)) for $Q=8$. Amazingly the frequency of this slow mode is precisely the fundamental frequency of this signal, and this modulation is the most intense for the highest harmonic ($f_6 = 6f_1$), this effect is due to the ordering of  these 6 frequencies as integer multiples of $f_1$, giving a constant frequency step between successive harmonics $f_{i+1} - f_i = f_1$. This $f_1= 1$ Hz slow modulation mode appears when two frequencies of the list are too close (in log-scale) for being separated properly by the analysing wavelet.

We observe  that in the time-frequency distribution of  $s$ shown in Fig.~\ref{sum_sin_model}(b) for $Q=8$, the higher harmonics cannot be distinguished, and their separation requires increasing markedly the value of $Q$ (for instance $Q=128$ in Fig.~\ref{sum_sin_model}(c)).  Fig.~\ref{sum_sin_model}(d)  illustrates two  sections of these distributions for $t=0$  (black curve, $Q=8$ and red curve, $Q=128$). This phenomena is even more visible on the ratio distribution $R[\mathcal{S}^{(Q)},\mathcal{S}^{(Q)}] (\log q,t)$ in Fig.~\ref{sum_sin_model}(h). $R[\mathcal{S}^{(Q)},\mathcal{S}^{(Q)}] (\log q,t)$ presents   an odd number of peaks, it is symmetric around the central peak ($q=1$).  In that example, each of the sixth frequency components contributes to this central peak. 11 lateral peaks emerge for $q>1$, and accumulate closer to the central peak. The positions of these peaks correspond to all the possible distinguishable frequency ratios of the signal, and the amplitude of these peaks is proportional to the number of combinations of frequencies that produces a given ratio. To distinguish all the peaks in Fig.~\ref{sum_sin_model}(g,h) it was necessary to increase $Q$ to 128. The total number of frequency ratios (for $q>1$) is $\sum_{i=1}^{n-1} i = n(n-1)/2$ if $n>2$ , in this example it is equal to 15. When there is no redundancy in the frequency ratios, for instance if harmonic frequencies are prime multiples of the fundamental frequency, each frequency ratio occurs once in $R[\mathcal{S}^{(Q)},\mathcal{S}^{(Q)}] (\log q, t)$. 

\subsection{Physiological signals: voice recordings}

The voice signals reported in this manuscript were selected from the {\bf VO}ice-{\bf IC}ar-f{\bf ED}erico II (VOICED) database \cite{cesari_new_2018} recorded by the ``Institute of High Performance Computing and Networking of the National Research Council of Italy (ICAR-CNR)'' and the Hospital University of Naples ``Federico II'' during 2016 and 2017. This database can be downloaded from the PhysioNet website \cite{goldberger_physiobank_2000}. It has been proposed lately as a new element in research on automatic voice disorder detection and classification. Together with medical phonetic examinations of a set of 208 individuals, among which 73 male and 135 female, voice signals, proportional to a local sound emission intensity, were acquired for about 4-5~s and sampled at 8000~Hz at 32~bit, vocal folds were examined by laryngoscopy and two medical questionnaires were collected at the ambulatories of Phoniatrics and Videolaryngoscopy of the ``Federico II''  Hospital of the University of Naples or at the medical room of the ICAR-CNR. The protocol description is reported in \cite{cesari_new_2018}.
Dysphonia is a quite common voice disorder (1/3 of adults will suffer from it once in their lifetime), it may originate from a functional or organic alteration of the vocal 
apparatus and its mechanics and may not systematically be considered as pathologic \cite{le_huche_voix_2010-2,le_huche_voix_2010-1}. On the one side, laryngoscopy is an invasive technique that gives a direct view of the physical alterations of the vocal tract \cite{berke_laryngeal_1993}. On the other side, the analysis of the voice acoustic signal is not intrusive and, thanks to the improvement of signal analysis methods, it can nowadays be used to guide or assist the recognition of the origin of a suspected dysphonia.  Voice classification methods from voice recordings by the recognition and quantification of the voice timbre (or tone color) has rapidly attracted the interest of electronic and computer science engineers. Globally, one can classify these methods in three groups \cite{jouvet_performance_2017}, (i) the time-domain methods which use autocorrelation functions or their variants \cite{ross_average_1974,lowell_spectral_2011}  to search for  repeatability between a temporal waveform and its time lagged version, (ii) frequency domain methods which locate characteristic frequencies and conclude to a spectral ``coloraturas'' for the voice, these methods meet rapidly their limitations if the signal is not stationary, (iii) time-frequency domain techniques \cite{mallat_wavelet_1999,cohen_time-frequency_1995,huang_hilbert-huang_2014}, that we have chosen for this study. 

\begin{figure}
\begin{center}
\hspace*{-1.3cm}\includegraphics[scale=0.24]{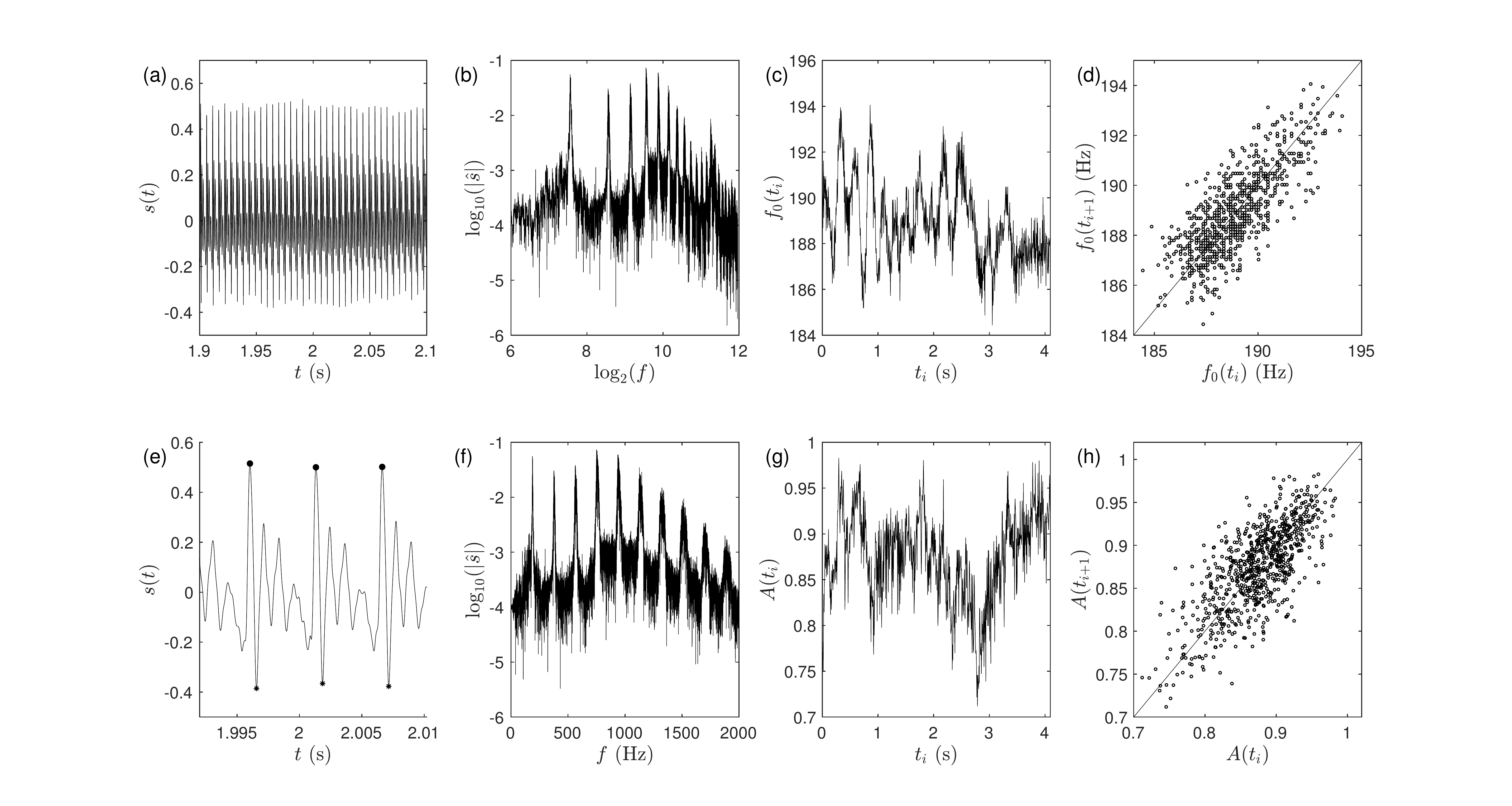}
\vspace{-0.5cm}
\caption{Analysis of the  \#008 voice signal $s(t)$ from the voice-icar-federico-ii database \cite{cesari_new_2018}. (a) Zoom of the signal during 0.2 s. (b) $|\hat{s}|$ plotted versus $f$ in logarithmic scales (base 10 and base 2 respectively). (c) Local frequency $f_1(t)$ computed from the detection of the extrema of larger prevalence from the signal (see (e)). (d) First return inverse of interpeak intervals $f_{1}(t_i)= 1/\Delta T_{i}$ scatter plot (these large amplitude peaks are marked with black dots in (e)). (e) Zoom of the signal on the short interval (20~ms) showing the local maxima $s_P(t_i)$ (black dots) and minima $s_p(t_i)$ (black stars) which are used to compute both the local frequency: $f_1(t_i) =1/\Delta T_{i}$ and the amplitude $A(t_i)$ of each larger amplitude peak: 
$A(t_i) = s_P(t_i) - s_p(t_i)$. (f) $|\hat{s}|$ (in base 10 log-scale) plotted versus $f$ in Hz (linear scale). (g) Amplitude of the largest peaks $A(t_i)$ versus time (see (e) for their detection). (h) First return peak amplitude ($A(t_{i+1})$ vs $A(t_i)$) scatter plot.  }
\label{voice008_fig1}
\vspace{-0.4cm}
\end{center} 
\end{figure}

The voice signal $s(t)$, numbered \#008, is that of a female of 51 years without deep vocal impairment  at the time of the test, ranked in the group of reflux laryngitis (Fig.~\ref{voice008_fig1}).  This example was chosen because it has marked peaks which can be detected by  thresholding the signal (this is quite rare, because it requires both a particular shape of the signal and a global stationarity of its amplitude). The Fourier spectra  of this signal (reported in log-log and log-lin scales in Fig.~\ref{voice008_fig1}(b) and (c) respectively) weight the power (in log-scale) of its spectral components; a fundamental mode with frequency $f_1 \sim 188$~Hz and higher modes (harmonics), ranked as integer multiples of $f_1$:  $i f_1$ with $i= 2, 3, 4, 5, ...$ with different powers. This simple frequency decomposition was observed in most of the signals provided in the VOICED database, this is a conspicuous characteristic of the human voice. These voice signals appear as the alternance of  quite regular large and sharp peaks (which give the fundamental mode) and smaller oscillations which may be very irregular. In some cases these smaller oscillations may be difficult to discriminate from the noise produced by some friction of the vocal tracts.  Even though this type of signal can be compared to the sum of sine functions introduced in Fig.~\ref{sum_sin_model}(e), the higher number of harmonics of this signal and their different power means that it  could be reproduced by a nonlinear dynamical system (ruled by nonlinear ordinary differential equations) where the different frequency components follow nonlinear rules \cite{travieso_detection_2017}. Our purpose in this paper is not to discuss the physical and biological mechanisms or the modelling  of  voice signals,  we have selected these examples as illustrations for our log-frequency correlation method because their spectral decomposition is very rich in harmonics (overtones) of the fundamental frequency. 

The temporal change of the fundamental mode frequency $f_1(t_i)$ and the largest peak amplitude $A(t_i)$ can be extracted from the \#008 voice signal 
 by thresholding its largest amplitude peaks (maxima: $s_P(t_i)$ and minima $s_p(t_i)$) as depicted in Fig.~\ref{voice008_fig1}(e). Fig.~\ref{voice008_fig1}(c) shows that $f_1(t)$ is modulated in time, suggesting an  irregularity of the rhythm coming from some difficulty of the patient to maintain a constant value of $f_1$. In this example, a similar temporal modulation is also visible on the largest peak amplitude $A(t_i)$ (Fig.~\ref{voice008_fig1}g). If these temporal variations were solely produced by instrumental  noise, the first return scatter plots of  $f_1$ and $A$ at successive peaks would give a symmetric cloud of points around the diagonal. In Fig.~\ref{voice008_fig1}(d) for the fundamental frequency modulation and in Fig.~\ref{voice008_fig1}(h) for the amplitude modulation these first return scatter plots are anisotropic, meaning that the dispersion of these values extends beyond instrumental noise. This conclusion is also confirmed by the temporal evolution of  $f_1(t_i)$ (Fig.~\ref{voice008_fig1}(c)) and $A(t_i)$ (Fig.~\ref{voice008_fig1}(g)), we notice that, in the first second, the modulations of $f_1(t_i)$ have the largest amplitude and are quasi-periodic, this first regime can also be recognised from the modulations of $A(t_i)$. This patient has a rather mild dysphonia  (classified as produced by reflux laryngitis) which can be recognised by an important set of harmonics  and a rather low vocal fold noise. 
  
\begin{figure}
\begin{center}
\hspace*{-0.3cm}\includegraphics[scale=0.235]{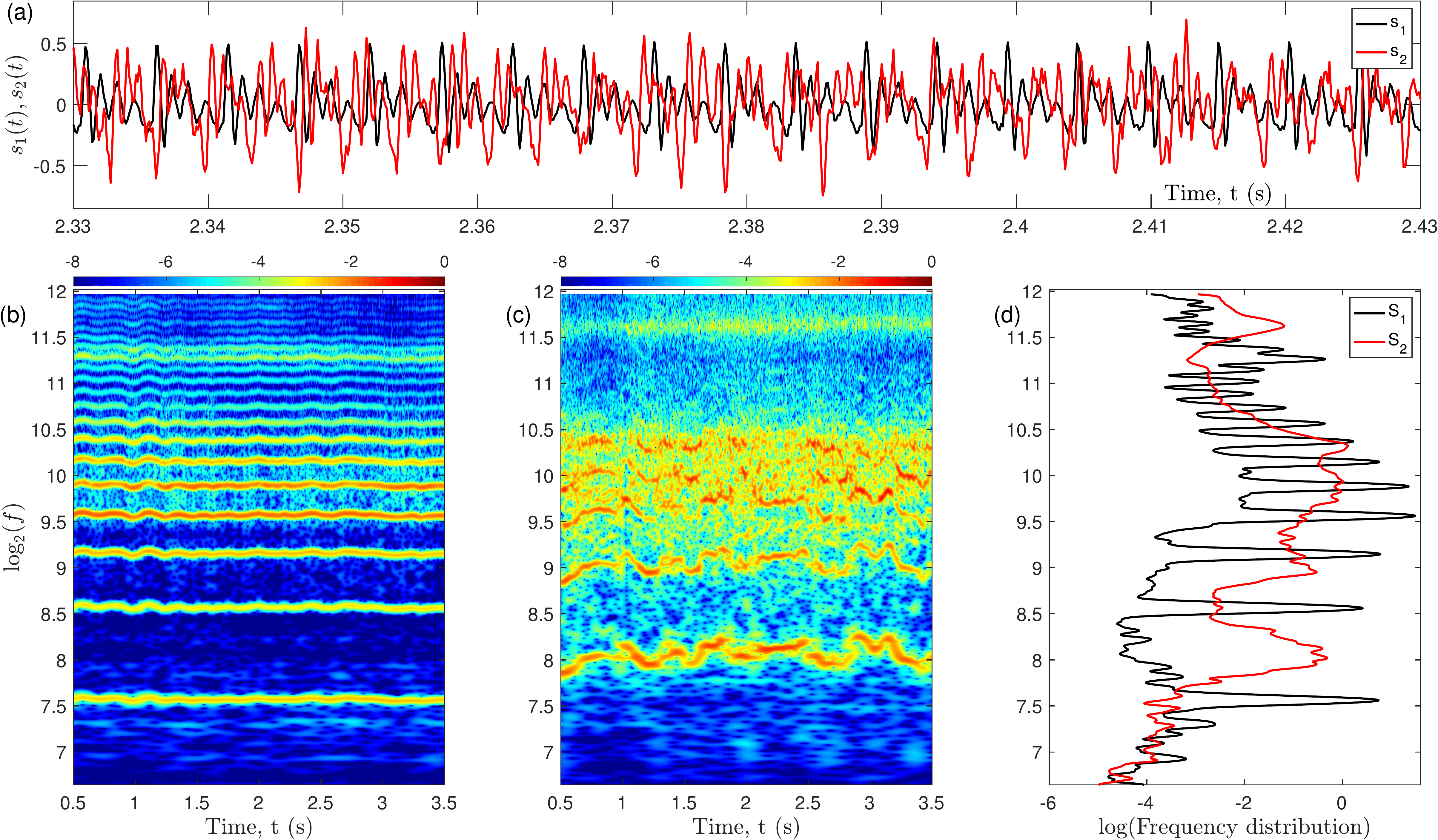}
\caption{Comparison of the time-frequency analysis of voice signals \#008 ($s_1$) and \#169 ($s_2$). (a) Zooms of $s_1$ and $s_2$ in a 0.1s window. (b,c) Associated time-frequency distributions (Eq.~\eqref{tfdist}) $\mathcal{S}^{(64)}[s_1](\log f,t)$, and $\mathcal{S}^{(64)}[s_2](\log f,t)$ computed with a Grossmann analysing wavelet and a quality factor $Q=64$.  
The  horizontal bands highlight the fundamental and harmonic frequencies. (d) Corresponding temporal averages of the frequency distributions reported in panels (b) black line and (c) red line. 
The ordinate of (d) (here the horizontal axes) is arbitrary and the frequency distributions are normalised. }
\label{Voices}
\vspace{-0.8cm}
\end{center} 
\end{figure}

 The second voice signal illustrated here is that of a female of 62~years (\#169 in the VOICED database), with hyperkinetic dysphonia. In that case, the quasiperiodicity observed in signal \#008  is so much disrupted that  it is impossible to use the previous threshold method for extracting the largest signal peaks;  the time-frequency analysis is required to check to which extent we can find a timbre for this voice, and how it changes with time. Fig.~\ref{Voices}(b,c) reports the colour-coded images of the \#008 ($s_1$) and the \#169 ($s_2$) time-frequency distributions $\mathcal{S}^{(64)}( \log f,t)$.  
With a temporal averaging of these time-frequency distributions $\langle\mathcal{S}^{(Q)}(\log f,t)\rangle_t$, we get a smooth estimate of power spectrum distributions for these two examples (Fig.~\ref{Voices}(d)). From the log-frequency filtering by the Grossmann analysing wavelet, the shapes of the averaged peaks shown in Fig.~\ref{Voices}(d) are in general different from those which could be obtained with a Welch estimator \cite{hayes1996}. The fundamental band frequency of \#169 is much broader than that of \#008, and shifted to greater values $f_{1,1} \sim 188.8$Hz (voice \#008) and $f_{1,2} \sim 268$Hz (voice \#169), and we also note  that it is quite impossible in \#169 to discriminate more than one harmonics from the averaged frequency spectrum. The time frequency distributions in Fig.~\ref{Voices}(b,c) highlight these differences. Whereas the fundamental mode band and its harmonics are weakly modulated in time for \#008, that of \#169 are very irregular, the third and fourth harmonics can be mixed up and indistinguishable, the harmonics above five are no longer visible.  The vocal folds of \#169 can no longer maintain their tight contact that is essentiel for a correct sound emission and the resulting effect, when hearing the voice, is that of a scratching noise which covers completely the expected tone. This person is quite unable to sing a melody.  

\subsection{Tuning voice  pitches via the computation of correlation functions}
\subsubsection{Reference frequency distribution $\mathcal{S}_{0,j}$}
For each signal $s_j$, an ``ideal'' frequency distribution $\mathcal{S}_{0,j}$ is first introduced in order to compare the real frequency distribution $\mathcal{S}^{(Q)}_j$ with a reference through the cross-correlation defined in Eq.~\eqref{spectrumrelations2}. Let us consider the vibrating string model used to represent the sounds emitted by stringed instruments. When the string is plucked at its ends, its natural frequencies are integer multiples of the fundamental frequency depending on the square root of the force of tension of the string \cite{geradin_mechanical_2015}. 
By analogy to this model, the reference frequency distribution $S_{0,j} (\log f)$ is a Dirac comb model in log scale defined as a sum of integer multiples of the fundamental frequency $f_{1,j}$ of the studied signal
\begin{align} \label{ref_fdist}
    \mathcal{S}_{0,j}(\log f)=\sum_n c_n \delta\left(\log \frac{f}{n f_{1,j}}\right) ,
\end{align}
with weights $c_n\geq0$ and possible cut-off ($c_n=0,\;\forall n>N$). We assume that the series of $c_n$ is bounded: $\sum_n c_n < +\infty$. 

We compare then the spectrum of the comb reference model to itself. We build an ideal ratio distribution by computing the auto-correlation of 
$\mathcal{S}_{0,j}$ derived from Eq.~\eqref{spectrumrelations3}:
\begin{align}\label{auto_corr_R00}
R_{00} (\log q) = R[\mathcal{S}_{0,j},\mathcal{S}_{0,j}] (\log q) = \sum_{n}\sum_{m} c_n c_m \delta \left(\log q\frac{n}{m}\right) ,
\end{align}
this ratio distribution  depends neither on time, nor on the signal under study.  When $\log q = 0$ ($q=1$), $R_{00}(0) = \sum_n c_n^2 < +\infty$. 

\subsubsection{Cross-correlation $R[\mathcal{S}_{0,j},\mathcal{S}^{(Q)}_j]$}
The time distribution cross-correlation function between $\mathcal{S}_{0,j}$ and $\mathcal{S}_{j}^{(Q)}$ is deduced from Eq.~\eqref{spectrumrelations3}:
\begin{align}\label{auto_corr_Rpsi}
R  [\mathcal{S}_{0,j},\mathcal{S}^{(Q)}_j] (\log q, t) = \sum_n c_n \mathcal{S}^{(Q)}_j (\log (q n f_{1,j}), t) .
\end{align}

Different degrees of ``frequency matching'' can also be captured by high peaks in $R[\mathcal{S}_0,\mathcal{S}^{(Q)}_j](\log q,t)$, especially when $q$ is a simple frequency ratio of harmonics of $f_{1,j}$ and $f_{1,0}$, for example 1:2 - octave, 2:3 - fifth, 3:4 - fourth would give a perfect consonance, 3:5 - major sixth and  4:5 - major third would give a medial consonance, 5:6 - minor third and 5:8 - minor sixth would give imperfect consonance. ``Unmatched'' frequency configurations would be obtained if a couple of frequency ratio of harmonics  belong to the dissonance list: 8:9 - major second, 8:15 - major seventh, 9:16 - minor seventh, 15:16 - minor second, 32:45 ($\sim 1/\sqrt{2}$) - tritone~\cite{helmholtz_theorie_1868}. 

In the following, we will denote $R_{ij} = R[\mathcal{S}^{(Q)}_i,\mathcal{S}^{(Q)}_j]$, the time-frequency window is fixed ($Q=64$). 

\subsubsection{Application to two voice signals from the VOICED data base}
For each of the two voice signals \#008 and \#169, we construct a Dirac comb model as reference distribution. These distributions are such that their lowest frequency peak matches the signal fundamental frequency (for instance for the signal \#008: $f_{1,1} = 188.8$Hz and for signal \#169: $f_{1,2} = 268$Hz).  The frequency of the highest harmonic of the comb model is limited by the sampling frequency $F_s$: $n f_{1,i} \lesssim F_s/2$  ($F_s = 8000$~Hz). We take $ c_n=1$, $\forall n \le15$: 
\begin{align} \label{ref_fdist_1}
    \mathcal{S}_{0,j}(\log f)=\sum_{n=1}^{15} \delta\left(\log \frac{f}{n f_{1,j}}\right) , \quad j=1,2.
\end{align}
For numerical computations, the frequency $f$ is discretized in $f_k = f_{\min}\cdot\alpha^{k-1}$,  with $k=1,2, 3, ...N$  and $N$ the size of the frequency vector $f$, $f_{\min}$ its minimum value and $\alpha$ the geometric factor determined from $N$, $f_{\min} =100$~Hz (fixed by the voice database), and $f_{\max} = F_s/2$. 
The correlation function $R_{0j} (\log q,t)$  is computed by combining this comb distribution with that of the voice signal as in Eq.~\eqref{Rcomp_numeric}, using the analytic expression of the $\log(f)$-Fourier transform ($\mathcal{F}_{\log f}$) of the comb distribution. We do not take its Fourier transform numerically because it is the source of numerical artefacts. For the comb model aligned to the fundamental frequency $f_{1,j}$ of signal $s_j$, it reads:
\begin{align}
\mathcal{F}_{\log f}\left[ \mathcal{S}_{0,j} \right] (u)= \sum_{n=1}^{15} \exp(-i 2\pi u \log (n f_{1,j}/f_{\min})) .
\end{align}
$u$ is the conjugated variable (through Fourier transformation) of $\log (f)$. In that space, we take the scalar product of $\mathcal{F}_{\log f}\left[ \mathcal{S}_{0,j} \right]$ with the conjugate of $\mathcal{F}_{\log f}\left[ \mathcal{S}^{(Q)}_{j} \right]$ (computed numerically from $W_{\psi_Q}[s_j](\log f,t)$) and compute its inverse Fourier transform to recover the correlation function $R[\mathcal{S}_{0,j},\mathcal{S}^{(Q)}_j](\log q)$.

\begin{figure}
\begin{center}
\includegraphics[scale=0.26]{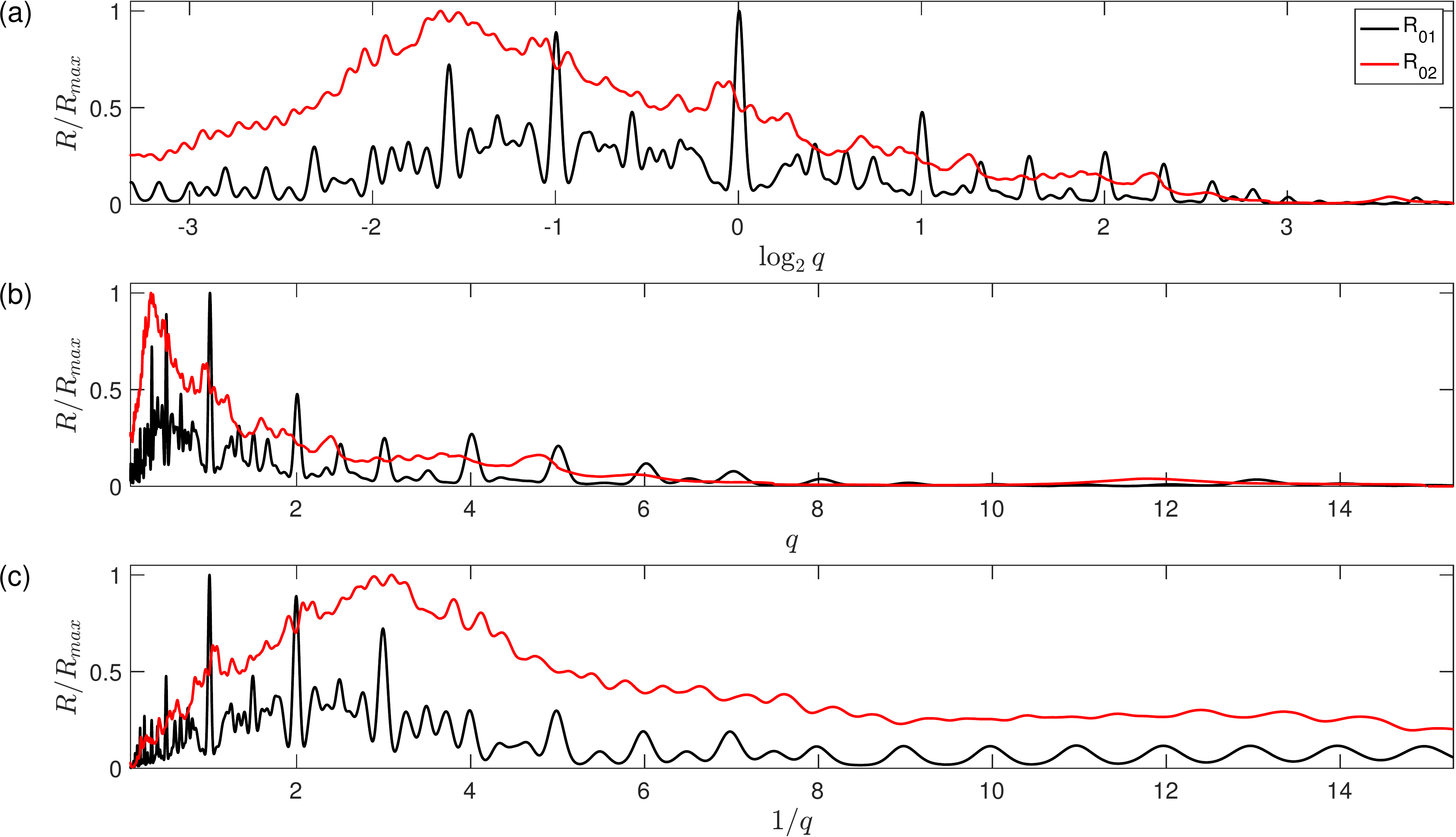} 
\caption{Ratio distributions of voice signals ((1) \#008 and  (2) \#169) and their frequency-matched Dirac comb models ($R_{0j}$). (a) Correlations of the frequency distribution $ \langle R_{0i}\rangle = R[\mathcal{S}_{0,j},\langle\mathcal{S}^{(Q)}_j\rangle_t](\log q)$  ($j=1$, 2, $Q=64$) with their reference comb frequency distributions (defined in the text). These correlations have been normalised to their maximum for ease of comparison. (b, c) Plot of $R[\mathcal{S}_{0,j},\langle\mathcal{S}^{(Q)}_j\rangle_t]$ versus $q$  for $q>1$ (b) and $1/q$ for  $q<1$ (c).  }
\label{Voices_relations}
\end{center} 
\vspace{-0.8cm}
\end{figure}

\begin{figure}
\begin{center}
\includegraphics[scale=0.29]{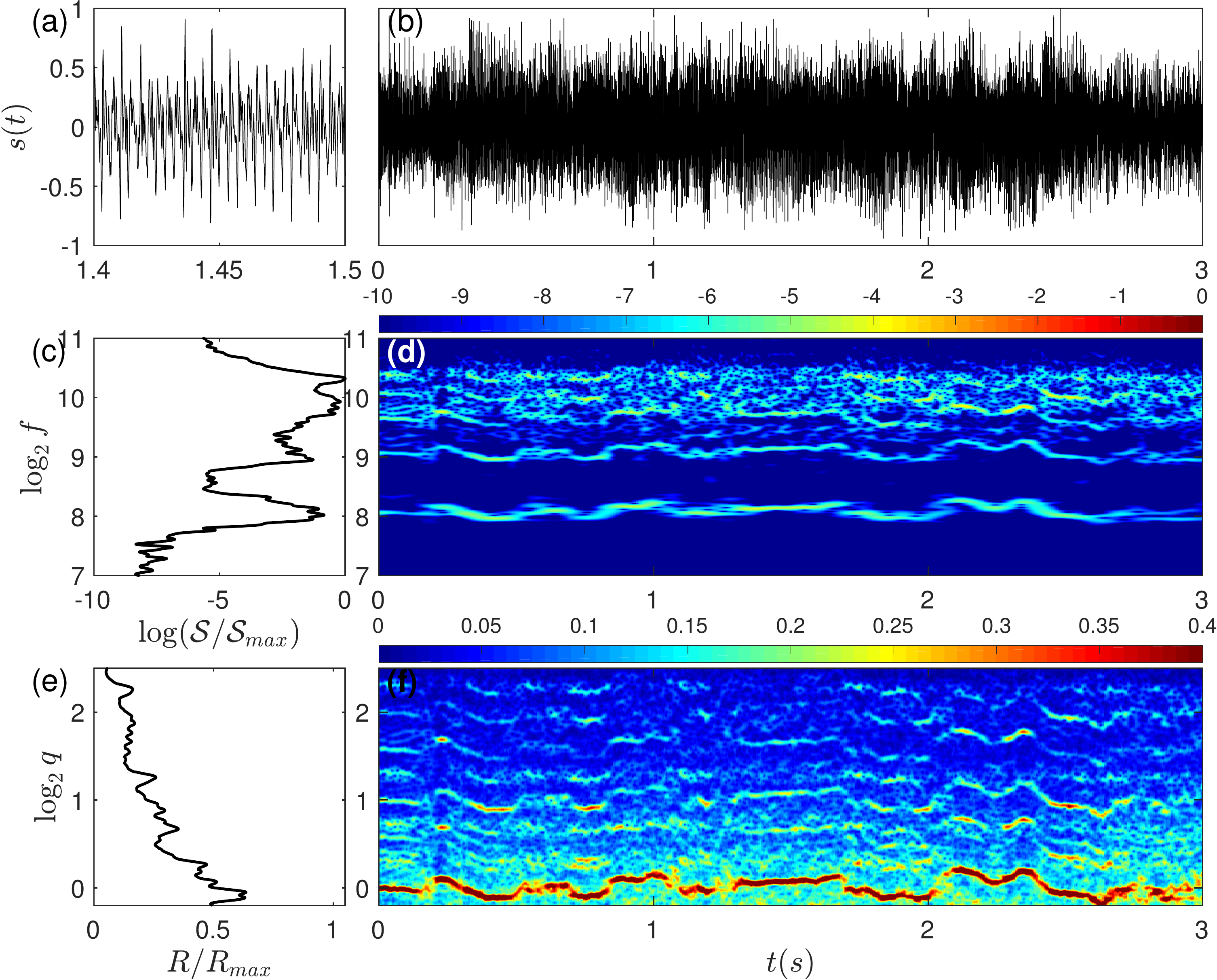} 
\caption{Comparing $R[\mathcal{S}_0,\mathcal{S}^{(Q)}] (\log q,t)$ with $\mathcal{S}^{(Q)} (\log f,t)$ for  voice signal \#169. (a)  Zoom of $s(t)$ in the [1.4s, 1.5s] interval. (b) Plot of a middle selection of $3$s out of the 4.5s recorded voice signal. (c) Temporal average of the frequency distribution  $\langle\mathcal{S}^{(Q)}\rangle_t (\log f) $ computed with a Grossmann analysing wavelet with quality factor $Q=64$.  (d) Colour-coded map of the time-frequency distribution  $\mathcal{S}^{(Q)} (\log f,t)  $. (e) Ratio distribution of the averaged frequency distribution: $R[\mathcal{S}_0,\langle\mathcal{S}^{(Q)}\rangle_t] (\log q)$.   (f) Colour-coded map of the time-ratio distribution  $R[\mathcal{S}_0,\mathcal{S}^{(Q)}] (\log q,t)$. }
\label{Voices_relations_tfreq}
\vspace{-0.3cm}\end{center} 
\end{figure}

The time-averaged frequency ratio distributions $R[\mathcal{S}_{0,j},\langle\mathcal{S}^{(Q)}_j\rangle_t] (\log q)$ of each voice signals \#008 and \#169 with its ``best-fitted'' Dirac comb model are presented in Fig.~\ref{Voices_relations}. If the signals were regular and quasi-stationary, these ratio distributions should pinpoint  ratios corresponding to the multiples of the fundamental frequencies. The plot of these correlation functions in linear $q$ and $1/q$ scales in Figs~\ref{Voices_relations}(b) and \ref{Voices_relations}(c)   highlights a strong asymmetry, it is due to different amplitudes of the fundamental mode and its harmonics compared to the constant coefficients in the comb model.  Again, as for frequency distributions, we note a strong difference of the ratio distributions for signals \#008 ($s_1$) and \#169 ($s_2$).  Confronting $R[\mathcal{S}_0,\mathcal{S}^{(Q)}] (\log q,t)$ with $\mathcal{S}^{(Q)} (\log f,t)$ for  voice signal \#169 (Fig.~\ref{Voices_relations_tfreq}) unveils important features which were not visible from the time averaged ratio distribution  $R[\mathcal{S}_0,\langle\mathcal{S}^{(Q)}\rangle_t] (\log q)$ (Fig.~\ref{Voices_relations}). Even if the fundamental mode frequency and its harmonics vary a lot during these 3s record, their ratios do not change dramatically, as a characteristic property of the mechanics of the vocal folds. In the middle of this signal ($1.4$s$<\! t \!< \!1.55$s) (see Fig.~\ref{Voices_relations_tfreq}(a) for a zoom in this interval), four flat ratio bands can be noticed, suggesting that this person put sufficient effort to recover for a short period of time a ``mild sensation'' of timbre. How this intermittent loss and recovery of the voice timbre occurs, the time range of these alternating sequences could be used as diagnosis criteria or aftercare follow-up (invasive intervention is necessary if soft or hard nodules are detected on the vocal cords (stage III), or voice exercises for earlier stages). It has been recently shown for patients with  neurodegenerative diseases, that not only the patient's ability to speak and formulate sentences is altered but also their voice purely acoustic features  \cite{al-hameed_new_2019}.  

\begin{figure}
\begin{center}
\hspace*{-0.28cm}\includegraphics[scale=0.22]{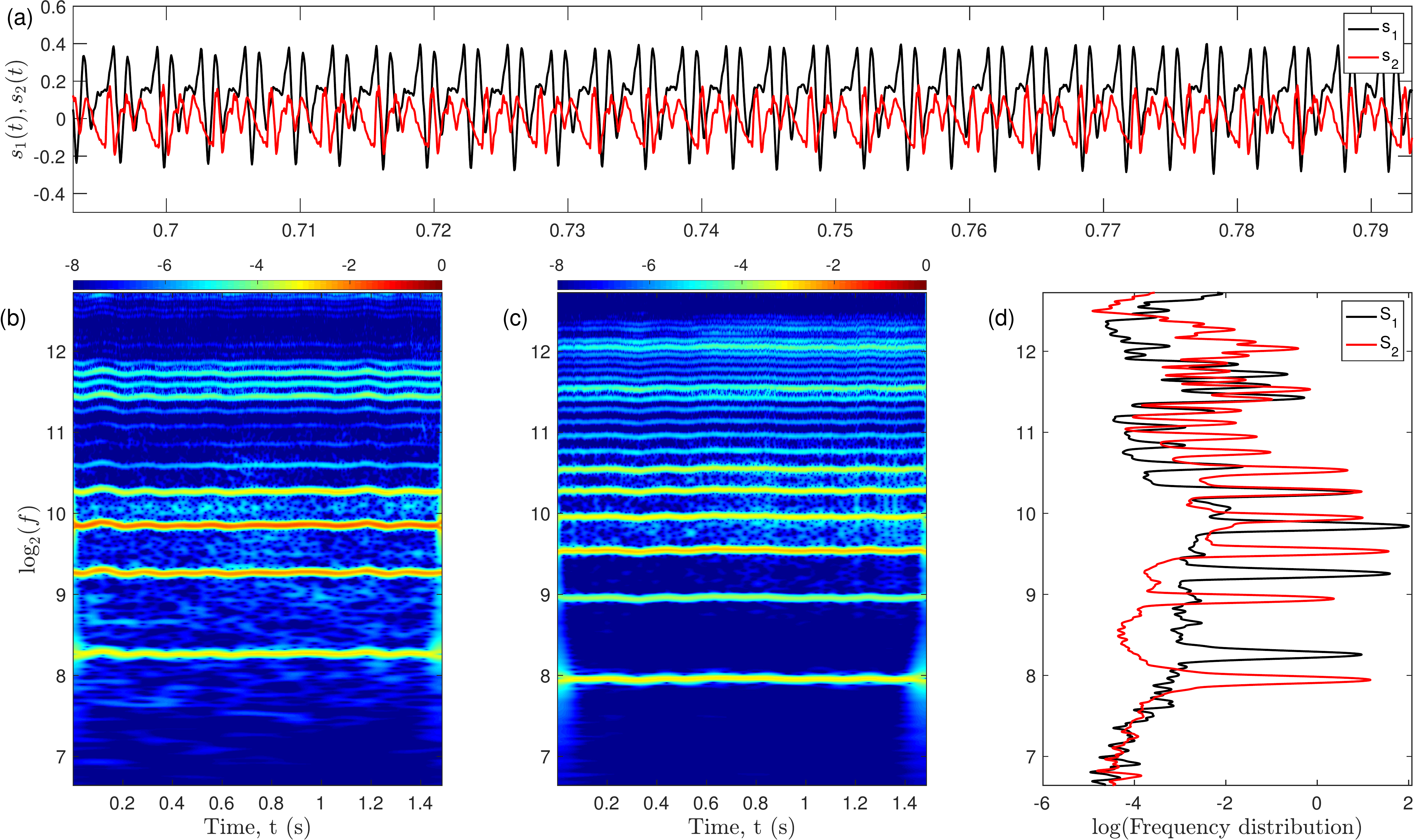}
\caption{Comparison of the time-frequency analysis  for two ``normal'' voice signals: $s_1$ is a sung vowel, $s_2$ is simply a maintained vowel. (a) Zooms of $s_1$ and $s_2$ in a 0.1s window. (b,c) Associated time-frequency distributions (Eq.~\eqref{tfdist}) $\mathcal{S}^{(Q)}_1(\log f,t)$, and $\mathcal{S}^{(Q)}_2(\log f,t)$ computed with a Grossmann analysing wavelet with quality factor $Q=64$.  
The  horizontal bands highlight the fundamental and harmonic frequencies. (d) Corresponding temporal averages of the frequency distributions reported in panels (b) black line and (c) red line. 
The ordinate of (d) (here the horizontal axes) is arbitrary and the frequency distributions are normalised. }
\label{Voices_singing}
\vspace{-0.8cm}\end{center} 
\end{figure}

\subsubsection{Cross-correlation $R[\mathcal{S}^{(Q)}_i,\mathcal{S}^{(Q)}_j]$ of two voice signals}
There are two interpretations for the log-frequency cross-correlation function $R[\mathcal{S}^{(Q)}_i,\mathcal{S}^{(Q)}_j] (\log q,t)$, leading to different possible applications. Either we see it as a distribution of the ratios $q$ between the frequencies of $\mathcal{S}^{(Q)}_i(\log f, t)$ and $\mathcal{S}^{(Q)}_j(\log f, t)$, or we view it for each $q$ as a measure of how well $\mathcal{S}^{(Q)}_i(\log f,t)$ and $\mathcal{S}^{(Q)}_j(\log (q f),t)$ match.
For instance, in the example reported here, $\mathcal{S}^{(Q)}_1,\mathcal{S}^{(Q)}_2$ are obtained from two different persons holding a pitch in their vocal range. 
The peaks in the correlation function $R[\mathcal{S}^{(Q)}_1,\mathcal{S}^{(Q)}_2](\log q)$ indicate the importance of the corresponding frequency ratios between the voices, in accordance with its first interpretation as a ratio distribution. When these ratios are close to simple rational numbers $m/n$, they indicate the presence of a $m:n$-synchronisation, that corresponds to the consonance of the voices simply sung together. This outlines our strategy to assess how rational the spectral relations of the voices are. The other interpretation is as follows: assuming $S^{(Q)}_2(\log qf)$ models the second voice transposed by $q$ to a different pitch, the peaks in $R[S^{(Q)}_1,S^{(Q)}_2](\log q)$ also indicate for which pitch transpositions the second voice would best match the first voice. This allows us to tune one voice with the other. 

\begin{figure}
\begin{center}
\includegraphics[scale=0.23]{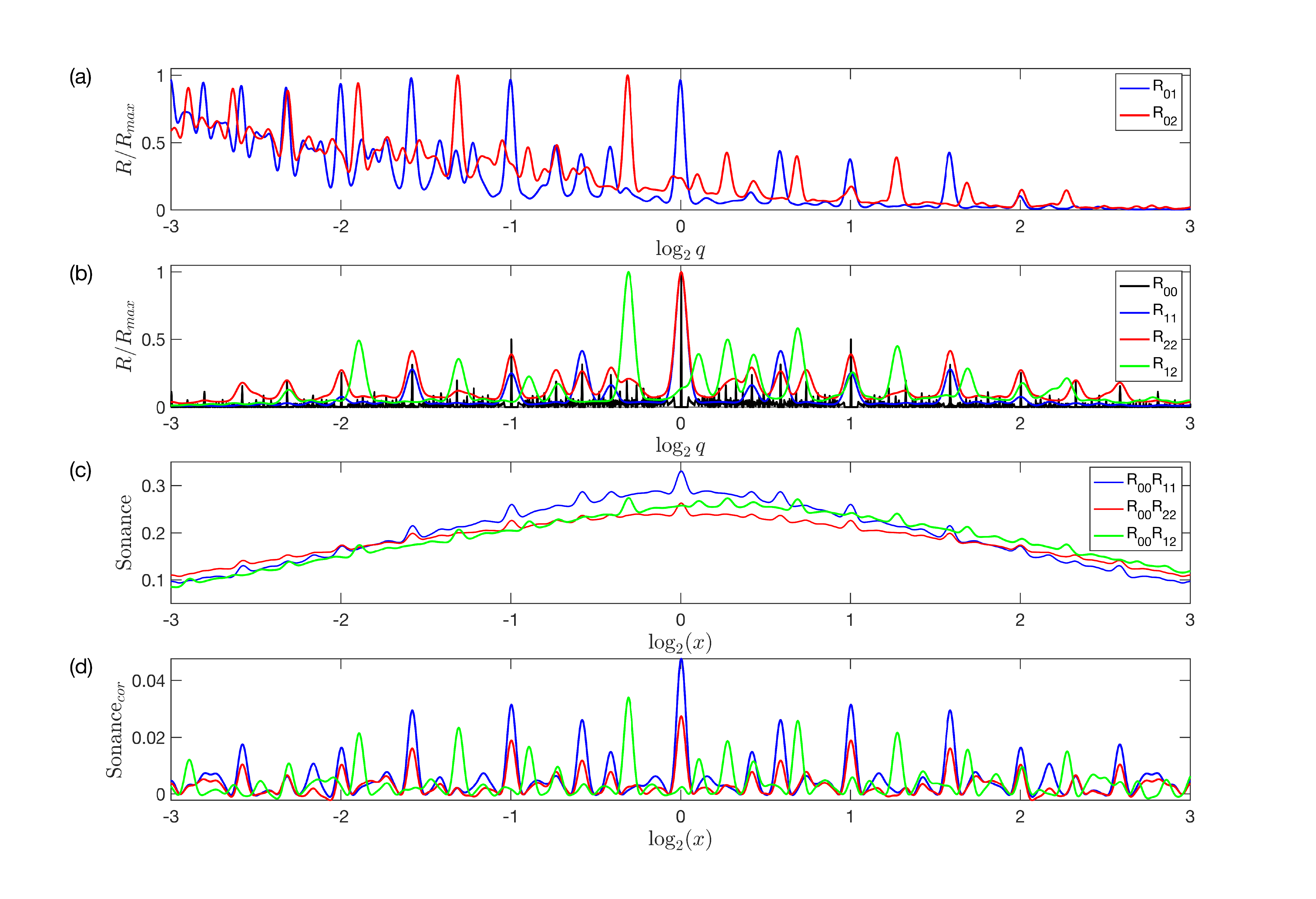} 
\vspace*{-0.2cm}\caption{Comparing ratio distributions and \emph{sonance} of  the two voice signals of Fig.~\ref{Voices_singing}. (a) Plots of the  normalised ratio distributions  $R_{0j}=R[\mathcal{S}_{0,1},\langle\mathcal{S}^{(Q)}_{j}\rangle](\log q) $ with $j=1,2$. For both signals we use the same Dirac comb model with the fundamental mode frequency  of voice $s_1$.  (b) Plots of the normalised ratio distributions $R_{11}$, $R_{22}$, $R_{12}$ with $R_{ij} = R[\langle\mathcal{S}^{(Q)}_{i}\rangle,\langle\mathcal{S}_{j}\rangle](\log q)$ computed from the two voices, and  $R_{00}=R[\mathcal{S}_{0,1},\mathcal{S}_{0,1}](\log q) $ computed from the comb Dirac model.  
(c) Plot of the \emph{sonance} (cross- and auto-) of the two voices with respect to the reference comb model ratio distribution for an arbitrary pitch transposition $\log (x)$: $\mus[R_{ij}](\log x) = R[R_{00},R_{ij}](\log x)$  with $i,j = 1,2$ green line, $i,j = 1,1$ blue line and $i,j=2,2$ red line.  
(d) Sonance curves of Fig.~\ref{Voices_relations_12}(c) are corrected by subtracting their lower envelope.    }
\label{Voices_relations_12}
\vspace{-0.8cm}\end{center} 
\end{figure}

The possibility to match a real voice signal  with reference model signals is very interesting because it can limit the maximum harmonics frequency for this cross-correlation, as an ``intelligent'' low pass filtering. This would not be possible by computing directly 
$R[\mathcal{S}^{(Q)}_i,\mathcal{S}^{(Q)}_j]$. 
We compare in Figs~\ref{Voices_singing} and \ref{Voices_relations_12} two different ``normal'' voices (``a'' vowel) from the clinic research in the speech therapy laboratory UNADREO  in Toulouse (France). 
The frequency distributions $\mathcal{S}^{(Q)}_{i} (\log f,t)$ and $\mathcal{S}^{(Q)}_{j} (\log f, t)$ ($i\ne j$) plotted in Fig~\ref{Voices_singing}, have the same characteristic frequency peaks structure as the voice \#008, but we notice that the frequency distribution of the voice $s_1$ has greater energy in the harmonics around 1000~Hz, which is characteristic of the emission of trained singer voices. 

The cross-correlation ratio distribution $R_{ij} = R[\mathcal{S}^{(Q)}_{i},\mathcal{S}^{(Q)}_{j}]$ is quite different from the auto-correlation ratio distributions $R_{00}$, $R^{(Q)}_{ii}$ and $R^{(Q)}_{jj}$. A common reference Dirac comb is chosen for both $s_1$ and $s_2$ ($i=1$ and $j=2$) and is aligned to the fundamental frequency $f_{1,1} = 307.2$Hz.  The highest central peak indicates the ratio of the fundamental frequencies, it is centered for $R_{00}$, $R_{11}$ and $R_{22}$ and shifted to $q = f_{1,2}/f_{1,1} =247/307.2 \sim 2^{-0.315}$ for $R_{12}$.  $R_{00}$ shows very sharp and narrow peaks which line up symmetrically  on either sides of $q=1$. The amplitude of these peaks recapitulates the weighting of the frequency ratios q for simple comb models and gives us which distribution would be obtained if all the frequency components of the signals had exactly the same power. We finish with the question of comparing the ratio distributions of the voices to the ideal reference. 

\subsection{Frequency distribution matching and \emph{sonance}}

We propose to find the best match between $R_{12}$ and $R_{00}$ by computing yet another cross-correlation:   
\begin{align}\label{sonance}
  &  R[R_{00},R_{12}](\log x,t) =\intoi R_{00}( \log q)R_{12}(\log x q, t)\d\log q \\ & =\intoi R_{00}\left( \log \frac{q}{x}\right)R_{12}(\log q, t)\d\log q =\sum_n \sum_m R_{12} \left( \log \frac{m}{n}x,t\right) .
\end{align}
This new quantity can be  obtained by two equivalent paths:
\begin{align}
    R[R_{00},R_{12}](\log x,t)= R[R_{01},R_{02}](\log x,t) ,
\end{align}
each corresponding to two uses of the newly introduced parameter $x$ that we call pitch transposition.
Either $R_{12}(\log x q,t)$ is seen as the distribution of ratios $q$ between the first voice $\mathcal{S}^{(Q)}_1(\log f,t)$ and the second voice of transposed pitch $\mathcal{S}^{(Q)}_2(\log x f,t)$. Or the $x$ in $R_{00}(\log(q/x))$ is seen as a varying ratio between the fundamental frequency of the ideal distribution $\mathcal{S}_0$.

Indeed, the best matching is expected when the pitch of the second voice is transposed to match the first one, thus when the voices are sung at unison, or equivalently when the fundamental frequency ratios are matched between the pairs of distributions: $x=f_{1,2}/f_{1,1}$.

As a result, for two voices of fundamental frequency $f_{1,1}$ and $f_{1,2} / x$, the quantity $R[R_{00},R_{12}](\log x,t)$ as a function of the pitch transposition $x$ has the following interpretation: it measures how ``ideal'' (similar to the model $R_{00}$) the spectral relations are between the voices. Extrema of this curve appear directly related to the musical property of consonance or dissonance of certain fundamental frequency ratios. For this reason, we call this quantity the \emph{sonance} between the two voices, and we rewrite its definition~\eqref{sonance} using the reference ratio distribution as the density of a measure $\d \mus(\log q)=R_{00}(\log q)\d\log(q)$: 
\begin{align}\label{sonance2}
   \mus[R_{12}](\log x,t) = R[R_{00},R_{12}](\log x,t) = \intoi R_{12}(\log x q,t)\d \mus(\log q) ,
\end{align}
that we could denote equivalently $\mus[\mathcal{S}^{(Q)}_1,\mathcal{S}^{(Q)}_2](\log x,t)$.

This \emph{sonance} measure is a geometric function of a pitch transposition quantity $x$, its maxima indicate the optimum relative pitch transpositions for which the two voices sung together would match best. 
 This term \emph{sonance} bears some analogy with  the concepts of consonance and dissonance which were first suggested by Pythagoras (sixth century BC), hence our choice of the symbol $ \mus$, but this similarity of terms must be nuanced. Dissonance and consonance are not mathematical quantities since they have been used to describe an empirical sensation  of human beings (combination of cochlea physiology and cognitive training) when hearing a mixture of sounds (two or more)\cite{plomp_tonal_1965,kameoka_consonance_1969}.
 
In Fig.~\ref{Voices_relations_12}, we compute the \emph{sonance} of the two voice records shown in Fig.~\ref{Voices_singing}, from time-averaged frequency distributions (for convenience) referenced to the same comb model. The comparison of the \emph{sonance} profiles in (c) to the ones of the generating auto- and cross-correlation functions $R_{01}$ and $R_{02}$ in (a), and $R_{12}$, $R_{11}$, $R_{22}$ and $R_{00}$ in (b) draws our attention to important features.  Apart from the asymmetry (symmetry) of the cross-(auto-)\emph{sonances, expected for any correlation function, the \emph{sonance} profiles have a strong positive baseline and are much less peaked than their simpler counterparts $R_{ij}$. This can be linked to the combined effects of a dense forest of peaks in $R_{00}$ and their width in the voices ratio distributions. Nevertheless, this baseline can be subtracted, as shown in Fig.~\ref{Voices_relations_12}(d), for further comparison. The highest peaks of $\mus[R_{ij}]$, pointing to the interval changes between the two voices that would lead to the best consonances, are very similar to the ones already present in $R_{ij}$. For instance, the global maxima of $\mus[R_{12}]$ to the left of $x=1$ corresponds to the ratio between the fundamental frequencies (voice $s_2$ has a lower pitch than voice $s_1$). However, their prominence are different, and \emph{sonance} profiles contain a more detailed landscape of smaller peaks and wells, tracking the gains and losses of consonance of the voices sung simultaneously at the corresponding relative pitch transposition. The almost inexistent} peak of $\mus[R_{12}](0)$ indicates that these two voices sung together without any tone adjustment would be quite dissonant because their fundamental and harmonic frequencies have few commensurability: the frequency ratios are not close to simple rational numbers $m/n$. In agreement with the common intuition, the central value of the auto-\emph{sonance} is also higher for the singer voice than for the untrained one, $\mus[R_{11}](0)>\mus[R_{22}](0)$.

The implementation of a frequency reassignment would be particularly beneficial to this time-dependent application, improving successively the discrimination of higher harmonics in the CWT, the resulting ratio distributions and the resolution of the \emph{sonance} landscape. In particular, recent developments in this area based on the synchrosqueezing transform~\cite{wu_instantaneous_2013, daubechies_conceft_2016} could be important for further developing the \emph{sonance} concept proposed herein. 

As a last remark, the \emph{sonance} profile of the voices is directly influenced by two choices: first, the quality factor of the wavelet $Q$, which determines the distinguishability of the frequencies and their ratios, and second, the choice of the number of harmonics and their amplitude in the reference comb model $\mathcal{S}_0$. We believe that, for a realistic \emph{sonance} profile, $Q$ should be related to the critical band of the ear~\cite{plomp_tonal_1965} and is, together with the design of the reference ratio distribution $R_{00}$, representative of the musical training of the ear.

\section{Conclusion}
We have introduced time-log-frequency ratio distributions based on analytic wavelets that we have applied to model and physiological signals (voice records). We found that the Grossmann wavelet is a natural shape for this task. A second correlation operation was defined to compare the matching of the voices ratio distribution with an ideally rational one, called \emph{sonance}. This function of a pitch transposition estimates, in a sense, the ``harmony'' produced by the two voices sung together. This work has shown that a geometric correlation function, in log-frequency is best suited to uncover characteristic frequency ratios between different signals. The application to voice records has been selected not only for its simplicity to perform and reproduce, but also because it gives credit to the concept of frequency ratios in voiced sounds. This method is presently generalized to physiological signals recorded from different organs or tissues, such as the heart and the breath, extending the application of these ratio distributions.

\begin{acknowledgement}
This work has been supported by ANR under contract: ANR-18-CE45-0012-01, 
EURLight S\&T funding.  We are very indebted to L. Delmarre, E. Harte,  and S. Polizzi for fruitful discussions. 
\end{acknowledgement}

\end{document}